\documentclass[a4paper,12pt]{article}
\usepackage[warn]{mathtext}				
\usepackage[T1, T2A]{fontenc}			
\usepackage[utf8]{inputenc} 			

\usepackage{verbatim}
\usepackage[utf8]{inputenc} 			 
\usepackage{wrapfig}				
\usepackage{amssymb, amsmath}	
\numberwithin{equation}{section}
\usepackage{indentfirst}   
\usepackage[titletoc]{appendix}
\usepackage{mathrsfs}
\usepackage{cmap} 	
\usepackage{bigints}
\usepackage[pdftex]{graphicx}
\usepackage{pgfplotstable}	
\usepackage[bottom]{footmisc}
\usepackage{hyperref}
\hypersetup{
     colorlinks   = true,
     linkcolor = blue,
     citecolor    = blue
}
\newcommand*{\defeq}{\stackrel{\text{def}}{=}}
\usepackage{graphicx}
\usepackage{relsize}
\usepackage{euscript}

\usepackage{float}
\usepackage[left=20mm,right=20mm,top=20mm,bottom=20mm,bindingoffset=0mm]{geometry}
\def\edited#1{{\textcolor{black}{#1}}}
\let\DS     = \displaystyle

\iffalse    

\def\Vect#1{{\underline{#1\hspace{-0.5mm}}\hspace{0.5mm}}}
\else
\newcommand{\Vect}[1]{\boldsymbol{\mathrm{#1}}{}}

\title{Heating of a semi-infinite Hooke chain}
\author{Sergei D. Liazhkov$^{1,2,3}$}
\date{%
    $^1$Institute for Problems in Mechanical Engineering of the Russian Academy of Sciences, St. Petersburg, Russia\\
    $^2$Peter the Great Saint-Petersburg Polytechnic University, St. Petersburg, Russia\\%
    $^3$National Research University Higher School of Economics, St-Petersburg, Russia\\[4mm]
    \today
}

\begin{document}

\maketitle

\abstract{We investigate unsteady ballistic heat transport in a semi-infinite Hooke chain with a free end subjected to an arbitrary heat source. We \edited{derive an exact integral representation for the kinetic
temperature, together with a weak-dissipation lattice approximation
and two continuum-type descriptions.} The \edited{continualization} of the discrete solution \edited{for the kinetic temperature in the weakly dissipative limit is obtained in two representations. The first is a symmetrical continuum solution based on an even extension of the source with respect to the boundary. The second is a discrete-continuum solution derived by applying a
large-time asymptotic approximation to the fundamental solution at
the fronts of the incident and reflected thermal waves in an
    instantaneously perturbed conservative semi-infinite chain. This solution retains a contribution from the interaction of the incident and reflected waves, which is absent in the symmetrical continuum solution. }
Analysis of sudden point heat supply  
\edited{ indicates that, even in the
non-dissipative chain, the discrete solution for the kinetic temperature exhibits pronounced large-time
slowing near the free end, consistent with a local nonequilibrium
plateau, whereas the symmetrical continuum representation predicts unbounded logarithmic growth.}
The results are expected to provide \edited{a benchmark for continuum models of heat transport in lattices with a free boundary}.

}

\section{Introduction}\label{sect1}
In this paper, we examine ballistic heat transport in a one-dimensional discrete medium. This regime occurs at the nanoscale, where quasiparticles (phonons) do not interact with each other. Experimental observations reveal features that violate \edited{the} Fourier law and are characteristic of ballistic transport, including direct proportionality between thermal conductivity and sample size\footnote{This is the limiting case of a power-law dependence of thermal conductivity on sample size, where the heat transport regime is referred to as either anomalous~\cite{Dhar2008} or quasi-ballistic \cite{Anufriev2018, Anufriev2024}. The other limiting case is the independence of thermal conductivity from sample size, corresponding to the diffusive regime.} (see, e.g., \cite{Chang, Vakulov, Maire, Cao, Luck}), independence of thermal resistance from sample size (see, e.g., \cite{Chang}), oscillatory decay of sinusoidal temperature profiles~(see, e.g.,~\cite{Minnich, huberman2019, ZH}), and wavelike heat propagation in solids (particularly in dielectrics, where heat travels via elastic waves, as demonstrated experimentally~\cite{Gutfeld, McNelly, Northrop2}).  Developing a comprehensive theory to describe these experiments and (micro-)nanoscale heat transport is essential for applications in (micro-)nanoelectronics \cite{PopSi, Cahil, Li, Malik, HAMR}, nuclear energy~\cite{Nuclear0}, and other fields discussed in~\cite{OTH}.
\par We focus on thermal transport in one-dimensional chains subjected to external heat supply and environmental heat exchange. Most theoretical studies describe the latter in a nonequilibrium steady state. Probably the first such work is~\cite{Rieder}, in which the temperature profile in the chain with the heat reservoirs is shown to be essentially constant within the chain (except at its fixed ends); \edited{for a
finite harmonic chain coupled to two Langevin reservoirs, the kinetic temperature is essentially the
arithmetic mean of the two bath temperatures (apart from deviations near
the contacts), and the heat current remains finite as the system size
increases, implying ballistic transport rather than the Fourier law.} A similar problem for a chain with free ends was addressed in~\cite{Nakazawa}. Subsequent research has incorporated disorder~\cite{Lebowitz1971}, anharmonicity \cite{Lepri1997, Kannan2012}, particle collisions \cite{Lepri2008}, spatial inhomogeneities \cite{Gend_2021_1, Gend_2021_2} and magnetic fields \cite{Dhar_Magn_1, Dhar_Magn_2}. Despite advances in steady-state problems, investigating non-stationary problems is required for a \edited{complete} understanding of heat propagation~(in particular, \edited{to determine the heat-transport regime}).
\par In~\cite{Gavr2018}, unsteady ballistic heat transport in a one-dimensional monoatomic harmonic lattice (i.e., in the infinite Hooke chain\footnote{The term <<Hooke chain>> was introduced in~\cite{ZAMM}.}) immersed in a viscous environment was studied. An analytical solution for the kinetic temperature was obtained using an approach that was later generalized in~\cite{Gavr2020}. Although \edited{the work}~\cite{Gavr2020} focused on steady-state kinetic temperature in an infinite square lattice, it derived an equation for unsteady heat propagation (see Eq.~(4.8) therein). Given the importance of describing thermal wave scattering from spatial inhomogeneities, analytical methods for investigation of unsteady heat propagation with boundaries are of interest.
\par Here, we study heat propagation in a one-dimensional semi-infinite Hooke chain with a free end subjected to external energy supply. \edited{The semi-infinite geometry is chosen in order to isolate the interaction of a thermal wave with a single free boundary and to avoid recurrent reflections from a second end.} \edited{The current work} extends~\cite{Sinf2023}, which \edited{considers an} instantaneous thermal pulse in the chain. We \edited{derive an exact integral representation for the kinetic temperature and, in the weakly dissipative limit, two continuum-type descriptions. The first is a symmetrical continuum solution based on an even extension of the source. As shown in~\cite{Sinf2023}, in the case of heat pulse this construction disagrees with the exact discrete and numerical solutions for the kinetic temperature at and near the free end. In the present paper, dealing with the supply, we show that in some cases this solution  even fails to capture the qualitative evolution of heat propagation along the chain. } Developing a new continuum approximation for the discrete solution remains an open challenge. Kinetic theory approaches face difficulties: the collisionless Boltzmann transport equation~(BTE) with evenness at the free end yields the mismatched continuum solution (see Remark 1 in \cite{Sinf2023}), while, \edited{to the best of our knowledge}, BTE with collision terms \edited{still} lacks effective \edited{treatment of} boundary scattering \edited{and of the associated} relaxation time. An alternative continualization route~\edited{(passage to the continuum limit)} is the quasi-continuum representation \edited{following} Kunin~\cite{Kunin, Charlotte2}, which successfully yielded continuum limits for ballistic heat transport in infinite Hooke chains~\cite{Sokolov2023}. However, applying this to semi-infinite chains may pose technical challenges. \edited{Therefore, in Sect.~\ref{S4_2}, we derive a new type of continuum solution, retaining a contribution from the interaction of the incident and reflected thermal waves~(interactions with the free boundary), which is absent in the symmetrical continuum solution.}
\par The paper is organized as follows. Section~\ref{sect_1_5} defines the notation. Section \ref{sect2} formulates the stochastic \edited{lattice} problem, following \cite{Gavr2018}. Section~\ref{sect3} \edited{derives an
exact integral representation for the kinetic temperature and its
weakly dissipative lattice reduction.} Section~\ref{sect4} \edited{presents two
continuum-type descriptions of this lattice solution}: one \edited{is} through the principle of symmetry formulated in \cite{Sinf2023} (Section~\ref{S4_1}) and the other \edited{is obtained} via the approach in Section~\ref{S4_2}. Section~\ref{sect_4_4} examines sudden heat supply, deriving integral-form continuum solutions for kinetic temperature and far-field asymptotics for supply at any point, considering both non-dissipative~(Section ~\ref{sect_4_4_1}) and dissipative~(Section ~\ref{sect_4_4_2}) cases. In the non-dissipative case of constant-intensity sudden heat supply, we \edited{observe an unexpected large-time slowing of the kinetic-temperature
growth. For an interior source, this effect is most pronounced at the
free end; when the source is placed at the free end, the discrete
temperature field at the heated sites is consistent with a gradual
approach to an apparent local plateau.}
Section~\ref{sect_6} discusses the results, \edited{limitations and possible extensions}.

\section{Nomenclature}\label{sect_1_5}
\par $\mathbb{N}$ is the set of all natural numbers;
\\$t$ is the time;
\\$n$ is a particle number;
\\$\dot{(...)}$ and~$\ddot{(...)}$ stand for the first and second time derivatives respectively;
\\$H$ is the Heaviside function;
\\ $\delta_{n,n'}$ is the Kronecker delta~($1$ if~$n=n'$ and~$0$ otherwise);
\\$\delta$~is the Dirac delta function;
\\$\langle ... \rangle$ stands for the expected value sign;
\\$\theta$ is the wave number;
\\$\omega_e$ is the elementary atomic frequency;
\\$x$ is a continuum spatial coordinate;
\\$a$ is the undeformed bond length;
\\$J$ is the Bessel function of the first kind;
\\$K$ is the modified Bessel function of the second kind.

\section{Statement of the problem}\label{sect2}
\begin{figure}[htb]
\center{\includegraphics[width=0.7\linewidth]{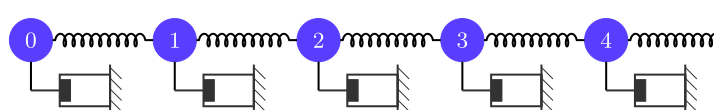}}
\caption{Semi-infinite chain of particles~(numbered), connected via the springs. The chain is immersed in a viscous environment.}
\label{fig_chain_damp}
\end{figure}

We consider the semi-infinite Hooke chain, which has one free end and is surrounded by a weakly viscous environment~(Fig.~\ref{fig_chain_damp}). 

We assume that the motion of particles is triggered by uncorrelated white noise. Therefore, the dynamics of particle~$n$ is governed by the following system of Langevin equations:
\begin{equation}\label{eq1}
    \begin{array}{l}
        \mathrm{d}u_n=v_n\mathrm{d}t,\\[2mm]

        \mathrm{d}v_n=\left(\omega_e^2(u_{n+1}-u_n)-\omega_e^2(u_n-u_{n-1})(1-\delta_{n,0})-2\eta v_n\right)\mathrm{d}t+b_n\mathrm{d}W_n, \quad n \in \mathbb{N}\cup\{0\},\\[2mm]
        \omega_e\defeq \sqrt{c/m},\quad \mathrm{d}W_n=\rho_n\sqrt{\mathrm{d}t},\quad \langle \rho_n \rangle=0,\quad \langle \rho_n(t_l) \rho_{n'}(t_{l'}) \rangle =\delta_{n,n'}\delta(t_l-t_{l'}),
        \end{array}
\end{equation}
where~$u_n$ and $v_n$~are displacement and velocity for $n$-th particle respectively;~$c$ is the bond stiffness; $m$~is the mass of particle; $\omega_e$ is the elementary atomic frequency;~$\eta$ is a viscosity of the environment; $W_n$~are uncorrelated Wiener processes;~$\rho_n(t)$ are independent random variables with zero mean and unit variance;~$b_n(t)$~is an intensity of the stochastic excitation. 
\par \small \textit{Remark}~1. 
The form of Eqs.~(\ref{eq1}), which is given also in~\cite{Gavr2018, Gavr2020} aside from the current work, is not the only form of the Langevin equations, and we use it because it seems more convenient to obtain the analytical solution. \edited{An equivalent formulation}~(where the right-hand side contains the term corresponding to white noise, that is, $\dot{W}$) \edited{can be found}, in particular, in~\cite{Nakazawa, Hemmer, Gavr2021}. \edited{In the present formulation, as in the heat-supply models
of~\cite{Gavr2018,Gavr2020,Gavr2022}, no
fluctuation--dissipation relation is imposed between \(b_n(t)\) and
\(\eta\). The noise is a prescribed mechanism of energy supply, not an
equilibrium thermal bath. This differs from Langevin-reservoir models
such as~\cite{Nakazawa}, where the noise intensity and the damping are
related by the bath temperature.}

\par \small \textit{Remark}~2. 
The \edited{structure of the semi-infinite} chain considered in the present paper may be studied not only in the context of lattice dynamics theory but also \edited{can be viewed as a reduced one-dimensional model of axial or torsional wave propagation in long engineering systems, including drill strings and pipelines} in petroleum engineering~\cite{Mirz}. \edited{In that
analogy the terminal particle may represent a free bit, and the present
stochastic forcing can be interpreted as a schematic counterpart of irregular external loading\footnote{This analogy was proposed by M.M. Khasanov in private communication.}.
}
\normalsize

Initially, all particles have zero displacements and velocities:
\begin{equation}\label{eq2}
u_n(0)=0,\qquad v_n(0)=0,
\end{equation}
i.e., dynamics of the particles induced by the stochastic noise only. \edited{In the next section, the statement problem~(\ref{eq1}---\ref{eq2}) is used to derive an exact integral representation for the kinetic
temperature.}

\section{An exact equation for the kinetic temperature}\label{sect3}

In this section, we derive an exact solution for the kinetic temperature. We consider an infinite set of realizations, which differ only by stochastic forces, related to uncorrelated Wiener processes. Due to the simplicity of the considered model~\edited{(the Hooke chain)}, we can characterize a thermal state via one parameter, which is referred to as a kinetic temperature:
\begin{equation}\label{eq4}
   k_\mathrm{B} T_n \defeq m\langle v_n^2\rangle.
   \end{equation}
There are several approaches to obtain the kinetic temperature, obeying the definition~(\ref{eq4}). The covariance approach, namely, the introduction of all types of covariance of particle displacements and velocities and the derivation of deterministic equations with respect to these covariances, is considered in~\cite{Gavr2018, Gavr2020} in the framework of the heat supply problem. The approach, based on the formulation of the Green function, is studied in~\cite{Kim}. An approach, which we use, is as follows. \par  To solve Eqs.~(\ref{eq1}), we use the direct and inverse discrete cosine transforms~(DCT):
\begin{equation}\label{DCT}
    \DS \hat{u}(\theta)=\sum_{n=0}^\infty u_n\cos \frac{(2n+1)\theta}{2},\quad u_n=\frac{1}{\pi}\int_{-\pi}^\pi \hat{u}(\theta)\cos \frac{(2n+1)\theta}{2}\mathrm{d}\theta,
\end{equation}
where~$\theta$ is the wave number; $\hat{u}$ is some time-dependent function. The solution for particle velocity is sought in the same way because~$\hat{v}(\theta)=\dot{\hat{u}}(\theta)$. The expression for~$v_n$ is further employed to obtain the kinetic temperature.

Substitution of the formula~(\ref{DCT}) into the definition of the kinetic temperature~(\ref{eq4})  yields respectively:
\begin{equation}\label{eq5}
\begin{array}{l}
    \DS \frac{k_\mathrm{B}}{m} T_n=\frac{1}{\pi^2}\iint_{-\pi}^{\pi}\langle \hat{v}_1\hat{v}_2\rangle \cos{\frac{(2n+1)\theta_1}{2}}\cos{\frac{(2n+1)\theta_2}{2}}\mathrm{d}\theta_1 \mathrm{d}\theta_2,\quad \hat{[...]}_i=\hat{[...]}(\theta_i),\\[3mm]
    \end{array}
\end{equation}
Therefore, to obtain the kinetic temperature, the covariance~$\langle \hat{v}_1 \hat{v}_2\rangle$ is yet to be obtained. Similarly, the problems of heat flux in the chain with reservoirs~\cite{Guzev1,Guzev2} and energy growth in the Hooke chain~\cite{Lykov} were solved.  The next section is dedicated to the solution with respect to the kinetic temperature in the semi-infinite chain.

\subsection{The image of particle displacements and velocities}

Applying~(\ref{DCT}) to the dynamic equations~(\ref{eq1}) with initial conditions~(\ref{eq2}) yields the following system with respect to~$\hat{u}$ and~$\hat{v}$:
\begin{equation}\label{eq7}
    \begin{array}{l}
   \DS \mathrm{d}\hat{u}=\hat{v}\mathrm{d}t,\\[2mm]
    \DS\mathrm{d}\hat{v}=\left(-\omega^2\hat{u}-2\eta \hat{v}\right)\mathrm{d}t+\sum_{n=0}^{\infty}b_n\mathrm{d}W_n\cos{\frac{(2n+1)\theta}{2}},
    \end{array}
    \end{equation}
where~$\omega(\theta) \defeq 2\omega_e \Big \vert\sin{\DS \frac{\theta}{2}} \Big \vert$ is the dispersion relation for the Hooke chain, with initial conditions
\begin{equation}\label{eq8}
    \hat{u}=\hat{v}=0,
\end{equation}
We introduce a vector~$\Vect{\hat{u}}(\theta)$ such that
\begin{equation}\label{eq9}
    \hat{\Vect{u}}(\theta)=\begin{pmatrix}\hat{u}(\theta) \\ \hat{v}(\theta) \end{pmatrix}.
\end{equation}
The equations~(\ref{eq7}) are rewritten into the first-order stochastic differential equation in the matrix form:
\begin{equation}\label{eq10}
    \mathrm{d}\hat{\Vect{u}}=\Vect{A}\hat{\Vect{u}} \mathrm{d}t+\Vect{B}\mathrm{d}\Vect{W},
\end{equation}
\begin{equation}\label{eq10_1}
\Vect{A}(\theta)=\begin{pmatrix}0 &  1 \\ -\omega^2(\theta) &  -2\eta\end{pmatrix},\quad \Vect{B}(\theta)=\begin{pmatrix}0 & 0 & ... \\ b_0\cos{\DS\frac{\theta}{2}} & b_1\cos{\DS\frac{3\theta}{2}} & ...\end{pmatrix},
\end{equation}
\begin{equation}\label{eq10_2}
\mathrm{d}\Vect{W}=\begin{pmatrix}\mathrm{d}W_0 \\\mathrm{d}W_1\\ ...\end{pmatrix},
\end{equation}

with initial condition
\begin{equation}\label{eq11}
\hat{\Vect{u}}=\Vect{0}.
\end{equation}

The following solution for~$\hat{\Vect{u}}$ is obtained in the form of the It{\^o} integral~(see Appendix~\ref{AAA}):
\begin{equation}\label{eq12}
    \hat{\Vect{u}}=\int_0^t e^{\Vect{A}(t-\tau)}\Vect{B}(\tau) \mathrm{d}\Vect{W}.
\end{equation}
Furthermore, we employ the obtained solution for~$\hat{\Vect{u}}$ to find a variance matrix, introduced in the next paragraph.

\subsection{The variance matrix} 
\edited{The second moments of linear Langevin dynamics driven by white noise
obey a closed equation. This fact is classical for the Brownian motion
of a single particle~\cite{uhlenbeck1930theory,chandrasekhar1943stochastic} and
was formulated systematically for the correlation functions of
linear stochastic systems driven by Gaussian white noise in~\cite{wang1945theory}. For harmonic chains, covariances of the
particle displacements and velocities have been used both in the
stationary two-bath problem~\cite{Rieder} and in the heat-supply
setting~\cite{Gavr2018,Gavr2020}; related displacement--velocity
correlations also determine the heat flux in finite
chains~\cite{Guzev1,Guzev2}.
Here we work with the covariance of the images of the displacement and velocity.} We introduce the matrix~$\Vect{D}$ such that
\begin{equation}\label{eq13}
    \Vect{D}\defeq\Bigg\langle \Bigl(\hat{\Vect{u}}_1-\langle \hat{\Vect{u}}_1\rangle\Bigr)\otimes\Bigl(\hat{\Vect{u}}_2-\langle \hat{\Vect{u}}_2\rangle\Bigr) \Bigg\rangle,
\end{equation}
where~$\otimes$ stands for the dyadic product of vectors\footnote{$\{\Vect{Q}\otimes\Vect{R}\}_{ij}=Q_iR_j$.}. Due to the uncorrelatedness of~$\rho$~(see~(\ref{eq1})), the mathematical expectation of~$\hat{\Vect{u}}$ equals zero. We rewrite~$\Vect{D}$ in the form

\begin{equation}\label{eq14}
    \Vect{D}=\Big \langle \hat{\Vect{u}}_1 \otimes \hat{\Vect{u}}_2\Big \rangle=\begin{pmatrix}\langle \hat{u}_1 \hat{u}_2\rangle & \langle\hat{u}_1 \hat{v}_2\rangle  \\ \langle\hat{v}_1 \hat{u}_2 \rangle &  \langle \hat{v}_1 \hat{v}_2 \rangle \end{pmatrix}.
\end{equation}
Therefore, the variance matrix $\Vect{D}$ is a covariance matrix, elements of which are covariances of images of velocities and displacements. On the other hand, $\Vect{D}$ can be calculated by substitution of the~$\hat{\Vect{u}}$~(\ref{eq12}) into~(\ref{eq14}) and simplification using uncorrelatedness of the Wiener processes. This yields~(see Appendix~\ref{BBB})

\begin{equation}\label{eq15}
    \Vect{D}=\int_0^t e^{\Vect{A}_1(t-\tau)}\left(\Vect{B}_1(\tau)\Vect{B}^\top_2(\tau)\right)e^{\Vect{A}_2^\top(t-\tau)}\mathrm{d}\tau,
\end{equation}
where~$\Vect{A}_i=\Vect{A}(\theta_i)$; $\Vect{B}_i=\Vect{B}(\theta_i)$;~$\top$ stands for the transpose sign. The sought covariance~$\langle \hat{v}_1 \hat{v}_2 \rangle$ is then found as~$D_{22}$. If necessary, the elements~$D_{12}$ and~$D_{21}$ may be used to obtain the heat flux, \edited{the expression for which contains covariances of the displacements and velocities~(see, e.g.,~\cite{Kuz2023, Guzev1, Guzev2}). The flux is an independent
quantity in the ballistic regime and remains for a separate study.}

\subsection{The discrete solution for the kinetic temperature in the semi-infinite chain}

Having multiplied matrices, containing in~the integrand of~$\Vect{D}$ subsequently and having taken~$D_{22}$, we obtain

\begin{equation}\label{eq16}
    \begin{array}{l}

   \DS \langle \hat{v}_1 \hat{v}_2\rangle=\mathlarger{\mathlarger{\sum}}_{n=0}^\infty
   b_n^2(t)* \Bigg(e^{-2\eta t}\prod_{i=1,2} \cos{\frac{(2n+1)\theta_i}{2}}\varphi_i(t)\Bigg), \\[3mm]
   
  \DS \varphi_i(t)=\cos{\left(t\sqrt{\omega(\theta_i)^2-\eta^2}\right)}-\frac{\eta\sin{\left(t\sqrt{\omega(\theta_i)^2-\eta^2}\right)}}{\sqrt{\omega(\theta_i)^2-\eta^2}}, \\[6mm]

  \DS X(t)*Y(t)\defeq\int_0^t X(\tau)Y(t-\tau)\mathrm{d}\tau=\int_0^t X(t-\tau)Y(\tau)\mathrm{d}\tau.
    \end{array}
\end{equation}

Following~\cite{Gavr2018}, introduce a function of heat supply intensity such that
\begin{equation}\label{eq17}
    k_\mathrm{B}\chi_n\defeq\frac{m}{2}b_n^2, \qquad \chi_n \big|_{t<0}=0.
\end{equation}
Substitution of~(\ref{eq16}) into~(\ref{eq5}), taking into account~(\ref{eq17}) and the property of multiplication of the integrals yields the following \textit{\edited{exact}} equation for the kinetic temperature:
\begin{equation}\label{KinTT}
\begin{array}{l}
\DS T_n=2\sum_{j=0}^\infty  \chi_j (t) * \left(e^{-2\eta t}\mathrm{I}_{n,j}^2(t) \right),\\[3mm]
\DS \mathrm{I}_{n,j}(t)\defeq\frac{1}{\pi}\int_{-\pi}^\pi \cos{\frac{(2j+1)\theta}{2}}\cos{\frac{(2n+1)\theta}{2}}\Bigg(\cos{\left(t\sqrt{\omega(\theta)^2-\eta^2}\right)}-\frac{\eta\sin{\left(t\sqrt{\omega(\theta)^2-\eta^2}\right)}}{\sqrt{\omega(\theta)^2-\eta^2}}\Bigg)\mathrm{d}\theta.
    \end{array}
\end{equation}
For~$\eta\ll \omega_e$, we simplify the latter, that is
\begin{equation}\label{KinTT2}
\begin{array}{l}
\DS T_n=2\sum_{j=0}^\infty  \chi_j (t) * \left(e^{-2\eta t}\dot{\Phi}^2_{n,j}(t)\right),\\[4.5mm]
\DS\dot{\Phi}_{n,j}(t)\defeq\frac{1}{\pi}\int_{-\pi}^\pi \cos{\frac{(2j+1)\theta}{2}}\cos{\frac{(2n+1)\theta}{2}}\,\cos({\omega(\theta)}t)\mathrm{d}\theta,
\end{array}
\end{equation}
where~\edited{the replacement of \(\mathrm{I}_{n,j}\) by \(\dot{\Phi}_{n,j}\), based
on \(\sqrt{\omega^2-\eta^2}\approx\omega\), has a
relative error of order \(O(\eta/\omega_e)\)}. \edited{Neglecting the latter}, we refer to~(\ref{KinTT2}) as the~\textit{discrete solution} for the kinetic temperature. Note that the second expression in~(\ref{KinTT2}) satisfies the following problem~\cite{Sinf2023}:
\begin{equation}\label{EQ25}
\ddot{\Phi}_n=\omega_e^2(\Phi_{n+1}-\Phi_n)-\omega_e^2(\Phi_n-\Phi_{n-1})(1-\delta_{n,0})+\delta_{n,j}\delta(t),
\end{equation}
corresponding to the dynamics of the semi-infinite chain after instantaneous perturbation of the particle~$j$ by the unit magnitude.  The equation~(\ref{EQ25}) is supplemented by the initial condition for~$\Phi_n$, written in the generalized form~\cite{Vladimirov}:
\begin{equation}\label{IC_PHI}
    {\Phi_n}|_{t<0}=0.
\end{equation}
Based on \edited{the above} and on the linearity of the problem, we write the structure of the discrete solution for the kinetic temperature in the Hooke chain with~\textit{arbitrary} boundary conditions:
\begin{equation}\label{KinTBC}
T_n=2\sum_{j\in \mathbb{P}}  \chi_j(t)*\left(e^{-2\eta t}\dot{\Phi}^2_{n,j}(t)\right),\quad \ddot{\Phi}_n=\omega_e^2\EuScript{L}_n\Phi_n+\delta_{n,j}\delta(t),
\end{equation}
where~$\EuScript{L}_n$ is the linear difference operator, which depends on specific boundary conditions; $\mathbb{P}$  is the set of numbers by which the perturbed particles in the system are indexed. Note that Eq.~(\ref{KinTBC}) is valid for~$\eta \ll \omega_e$ only. For \textit{arbitrary}~$\eta$, \edited{the corresponding exact Green function
representation is}:
\begin{equation}\label{KinTBC_Arb_eta}
T_n=2\sum_{j\in \mathbb{P}}  \chi_j(t)*\dot{\Phi}_{n,j}^2(t),\quad \ddot{\Phi}_n=\omega_e^2\EuScript{L}_n\Phi_n-2\eta \dot{\Phi}_n+\delta_{n,j}\delta(t),
\end{equation}
\edited{where \(\Phi_{n,j}\) is the fundamental solution of the damped lattice
problem stated on the right-hand side of the second equation in~(\ref{KinTBC_Arb_eta})}. In the next section, continualization procedure for the discrete solution for the kinetic temperature is discussed.

\section{Kinetic temperature in the continuum limit}\label{sect4}
In the case of an arbitrary heat supply source,~$\chi$, the formula~(\ref{KinTT2}) becomes hard to use. Therefore, \edited{we pass to continualized
descriptions of the kinetic temperature as a function of a continuous
coordinate~$x$. This representation not only simplifies the solution and physical interpretation but also bridges (micro-) nanoscopic and macroscopic heat transport descriptions.} 
\subsection{The first approach}\label{S4_1}
The first approach is based on using the continuum solution for the kinetic temperature in the \textit{infinite} chain, obtained in \cite{Gavr2018}:
\begin{equation}\label{ContT}
    T(t,x)=\iint_{-\infty}^{\infty}\chi(\tau,\xi) \mathcal{G}(t-\tau,x-\xi)\mathrm{d}\xi\mathrm{d}\tau,
\end{equation}
\begin{equation}\label{ContTelegraph}
    \mathcal{G}(t,x)\defeq\frac{H(v_st-|x|)e^{-2\eta t}}{\pi \sqrt{v_s^2t^2-x^2}},
\end{equation}
where~$v_s\defeq \omega_ea$ is the speed of sound; $\mathcal{G}(t,x)$ is the fundamental solution for the problem of heat transport in the infinite Hooke chain, immersed in a viscous environment~\cite{Gavr2018}; $T(t,x)$ and~$\chi(t,x)$ are respectively the continuum functions of the kinetic temperature and the heat supply intensity, such as
\begin{equation}\label{ContSource}
   T(t,an)= T_n(t),\qquad \chi(t,an)= \chi_n(t).
\end{equation}
We rewrite function~$\mathcal{G}$ as 
\begin{equation}\label{GCONT}
\mathcal{G}(t,x)=2e^{-2\eta t}G(t,|x|),\quad G(t,x)\defeq \frac{H(v_st-x)}{2\pi \sqrt{v_s^2t^2-x^2}},
\end{equation}
where~$G(t,|x|)$ is the well-known fundamental solution of the problem of ballistic heat transport in the Hooke chain, which was derived in the frameworks of both the lattice dynamics approach~\cite{Kriv2015, Sokolov2021, Allen, Kuz2021} and the kinetic theory~\cite{Kuz2021}. 
\par \small \edited{\textit{Remark}~3.  
The function~$G(t,|x|)$ coincides with the fundamental solution of the two-dimensional wave equation, where the second argument is the distance from a point source. In the Hooke chain, thermal energy, corresponding to~$G$, is equally distributed among the normal modes, each of which carries energy at its own group velocity. A localized heat pulse therefore behaves as if a point source in the plane emitted waves uniformly in all directions: one-dimensional dispersion mimics geometric spreading in a two-dimensional nondispersive medium\footnote{This analogy was proposed by Yu. V. Petrov in private communication.}.
}
\normalsize

Applying the principle of symmetry of the continuum solution~\cite{Sinf2023}, we rewrite the equation for the kinetic temperature in the semi-infinite chain:
\begin{equation}\label{TEMP_CLASSIC_CONT}
\begin{array}{l}

    \DS T(t,x)=\iint_{-\infty}^{\infty}\chi(\tau,|\xi|) \mathcal{G}(t-\tau,x-\xi)\mathrm{d}\xi\mathrm{d}\tau\\[3mm]\DS=
  2\int_0^\infty \int_\frac{|x-\xi|}{v_s}^t \chi(\tau,\xi)e^{-2\eta (t-\tau)}G(t-\tau,|x-\xi|)\mathrm{d}\tau\mathrm{d}\xi\\[3mm]
    +\DS2\int_0^\infty \int_\frac{x+\xi}{v_s}^t \chi(\tau,\xi)e^{-2\eta (t-\tau)}G(t-\tau,x+\xi)\mathrm{d}\tau\mathrm{d}\xi.
\end{array}
\end{equation}
We further refer to Eq.~(\ref{TEMP_CLASSIC_CONT}) as a~\textit{symmetrical continuum solution}.
\subsection{The second approach}\label{S4_2}
The second approach is based on the direct continualization procedure of the discrete solution~(\ref{KinTT2}). The continualization can be performed in two ways. The first, namely, the averaging of the discrete solution over a mesoscale, the order of which is more than~$a$ but less than the chain length, is described in detail in~\cite{Sinf2023, Kuz2021}. Applying it to the current problem obviously yields the symmetrical continuum solution, because the principle of symmetry of the continuum solution, formulated in~\cite{Sinf2023}, is the corollary of the method of continualization described above. The second way is presented below. 
\par In~\cite{Sinf2023} the problem~(Eqs.~(\ref{EQ25}),~(\ref{IC_PHI})) was shown to be equivalent to the problem for the infinite chain with the same source in the particle~$j$ and the mirrored source with respect to the center symmetry. In the framework of the discrete problem, it is \textit{not}~ particle~$0$ but rather \textit{the bond} between the~\edited{particles~$0$ and~$-1$.} Therefore, the problem governed by~Eq.~(\ref{EQ25}) is equivalent to the following one for the infinite chain:  
\begin{equation}\label{EQ26}
\ddot{\Phi}_n=\omega_e^2(\Phi_{n+1}-2\Phi_n+\Phi_{n-1})+(\delta_{n,j}+\delta_{n,-j-1})\delta(t),
\end{equation}
\edited{with initial condition~(\ref{IC_PHI}). In turn, the problem is equivalent to the 
following one}:
\begin{equation}\label{EQ226}
\ddot{\Phi}_n=\omega_e^2(\Phi_{n+1}-2\Phi_n+\Phi_{n-1}),
\end{equation}
with initial conditions
\begin{equation}\label{EQ226_IC}
\Phi_n=0,\qquad \dot{\Phi}_n=\delta_{n,j}+\delta_{n,-j-1},
\end{equation}
solution of which is written as follows~\cite{Hemmer, Schroed}: 
\begin{equation}\label{SOL_EQ26}
\begin{array}{l}
   \DS \dot{\Phi}_{n,j}=J_{2|n-j|}(2\omega_et)+J_{2|n+j+1|}(2\omega_et).\\[2mm]
\end{array}
\end{equation}
For the semi-infinite chain, the Eq.~(\ref{SOL_EQ26}) represents a superposition of the running wave, incident at the point~$j$, and reflected from the boundary wave~(described by the first and second terms, respectively). Furthermore, considering the semi-infinite chain, we omit the sign of the absolute value in the index of the second term in~Eq.~(\ref{SOL_EQ26}). 
\par We perform an asymptotic estimate of the function~$\dot{\Phi}$ \edited{at large times} at the fronts of incident and reflected waves, using the technique proposed in~\cite{Gavr2022}\footnote{For the first time~(for discrete media), this approach was used in~\cite{Gavr2022} to estimate the fundamental solution for the infinite Hooke chain.}, based on the stationary phase approximation. That is, the first and second terms in~(\ref{SOL_EQ26}) are estimated at the fronts of incident and reflected waves, respectively. Having done the latter, we write the estimate for~$\dot{\Phi}_{n,j}$ as a sum of the two \edited{quasiharmonics}, namely
\begin{equation}\label{ASYMP_SOL}
\dot{\Phi}_{n,j}\sim \check{\Phi}_{|n-j|}H\left(\omega_et-|n-j|\right)+\check{\Phi}_{n+j+1}H\left(\omega_et-(n+j+1)\right),
\end{equation}
\begin{equation}\label{ASYMP_SOL_2}
\check{\Phi}_n\defeq\sqrt{\frac{1}{\pi\sqrt{\omega_e^2t^2-n^2}}}\cos\left(\mathcal{W}_nt-\frac{\pi}{4}\right),\,\; \mathcal{W}_n(t)\defeq2\omega_e\left(\sqrt{1-\frac{n^2}{\omega_e^2t^2}}-\frac{n}{\omega_et}\arccos\frac{n}{\omega_et}\right),
\end{equation}
where~$\sim$ stands for the sign of the large-time asymptotically equivalent function; ~$\mathcal{W}_n$ is referred to as a characteristic frequency of the natural oscillations in the Hooke chain such that the frequencies of the incident and reflected waves are~$\mathcal{W}_{|n-j|}$ and~$\mathcal{W}_{n+j+1}$ respectively. 
Having analyzed the difference between the frequencies of the incident and reflected waves, $\Delta \mathcal{W}\defeq\mathcal{W}_{|n-j|}-\mathcal{W}_{n+j+1}$, we suppose it to  be negligible when~$j=0$ and when~$n=0$. In order to confirm the supposition, we put~$j=0$ and expand~$\Delta \mathcal{W}$ into series at~$1/\omega_et=0$. The latter yields 
\begin{equation}\label{DELTA_FREQ}
    \Delta \mathcal{W}\vert_{j=0}=\frac{\pi}{t}-\frac{2n+1}{\omega_et^2}+O\left(\omega_e^{-2}t^{-3}\right).
\end{equation}
The estimation for~$\Delta \mathcal{W}\vert_{n=0}$ satisfies Eq.~(\ref{DELTA_FREQ}) with~$n$ replaced by~$j$. Therefore, at large times~$\Delta \mathcal{W} \ll \omega_e$ at~$n=0~ \forall j$ and at~$j=0~\forall n$. This observation is important for asymptotic analysis of the expression for the thermal fundamental solution,~$\dot{\Phi}^2_{n,j}$.\par
In order to analyze the expression for the kinetic temperature~(Eq.~(\ref{KinTT2})) write separately asymptotic estimate for the~$\dot{\Phi}_{n,j}^2$:

\begin{equation}\label{SOLL}
\begin{array}{l}
\DS \dot{\Phi}^2_{n,j}\sim\dot{\mathit{\Phi}}^2_{n,j}\defeq\frac{1+\sin\left(2\mathcal{W}_{|n-j|}t\right)}{2\pi\sqrt{\omega_e^2t^2-(n-j)^2}}H\left(\omega_et-|n-j|\right)\\[4mm]\DS +\frac{1+\sin\left(2\mathcal{W}_{n+j+1}t\right)}{2\pi\sqrt{\omega_e^2t^2-(n+j+1)^2}}H\left(\omega_et-(n+j+1)\right)\\[4mm]
\DS+\frac{\sin\left(\left(\mathcal{W}_{|n-j|}+\mathcal{W}_{n+j+1}\right)t\right)+\cos\left(\left(\mathcal{W}_{|n-j|}-\mathcal{W}_{n+j+1}\right)t\right)}{\pi \sqrt[4]{\left(\omega_e^2t^2-\left(n-j\right)^2\right)\left(\omega_e^2t^2-\left(n+j+1\right)^2\right)}}H\left(\omega_et-(n+j+1)\right).
\end{array}
\end{equation}
We refer to the expression~(\ref{SOLL}) as a thermal fundamental solution. The first term in the latter is represented as a contribution of the \edited{incident} wave, and the second and third terms are contributions of the reflected wave. \edited{The remaining term is their cross contribution. By the product-to-sum
identity, this cross contribution contains quasi-harmonic components
with the sum and difference of the incident and reflected
characteristic frequencies. It therefore represents their interference and is responsible for the
corresponding contribution retained below.} Following~\cite{Gavr2022, Gavr2023_1}, we separate the thermal fundamental solution previously as
\begin{equation}\label{FASTSLOW_1}
    \begin{array}{l}
\DS\dot{\mathit{\Phi}}^2_{n,j}=\dot{\Tilde{\mathit{\Phi}}}^2_{n,j}+\dot{\hat{\mathit{\Phi}}}^2_{n,j},\\[3mm]
\end{array}
\end{equation}
\begin{equation}\label{FASTSLOW_2}
    \DS \dot{\Tilde{\mathit{\Phi}}}^2_{n,j}\defeq \frac{H\left(\omega_et-|n-j|\right)}{2\pi\sqrt{\omega_e^2t^2-(n-j)^2}}+\frac{H\left(\omega_et-(n+j+1)\right)}{2\pi\sqrt{\omega_e^2t^2-(n+j+1)^2}},
\end{equation}
\begin{equation}\label{FASTSLOW_3}
\begin{array}{l}
     \DS \dot{\hat{\mathit{\Phi}}}^2_{n,j}\defeq \frac{H\left(\omega_et-|n-j|\right)\sin\left(2\mathcal{W}_{|n-j|}t\right)}{2\pi\sqrt{\omega_e^2t^2-(n-j)^2}}+\frac{H\left(\omega_et-(n+j+1)\right)\sin\left(2\mathcal{W}_{n+j+1}t\right)}{2\pi\sqrt{\omega_e^2t^2-(n+j+1)^2}}\\[5mm]
  +\DS \frac{H\left(\omega_et-(n+j+1)\right)\sin\left(\left(\mathcal{W}_{|n-j|}+\mathcal{W}_{n+j+1}\right)t\right)}{\pi \sqrt[4]{\left(\omega_e^2t^2-\left(n-j\right)^2\right)\left(\omega_e^2t^2-\left(n+j+1\right)^2\right)}}\\\DS+\frac{H\left(\omega_et-(n+j+1)\right)\cos\left(\left(\mathcal{W}_{|n-j|}-\mathcal{W}_{n+j+1}\right)t\right)}{\pi \sqrt[4]{\left(\omega_e^2t^2-\left(n-j\right)^2\right)\left(\omega_e^2t^2-\left(n+j+1\right)^2\right)}},  
\end{array}
\end{equation}
where~$\dot{\Tilde{\mathit{\Phi}}}^2_{n,j}$ and~$\dot{\hat{\mathit{\Phi}}}^2_{n,j}$ are the slow and conditionally fast motion components respectively. The conditionally fast motion component represents the sum of four quasiharmonics. However, since the first two quasiharmonics correspond to the transient process, related to equilibration of the kinetic and potential energies, we neglect these. Since the characteristic frequencies of incident and reflected waves are close to each other when~$j=0~\forall n$ and for~$n=0~\forall j$~(see Eq.~(\ref{DELTA_FREQ})), the last quasiharmonic in Eq.~(\ref{FASTSLOW_3}) transforms to a slow quantity. We conclude that the perturbation at the boundary \textit{dramatically} changes the character of thermal wave propagation. A possible explanation of the latter \edited{is the partially destructive interference of the incident and
reflected thermal waves, which affects short-wavelength modes strongly. The corresponding
contribution is encoded in the third and fourth terms of~$\dot{\hat{\mathit{\Phi}}}_{n,j}^2$ (see Eq.~\eqref{FASTSLOW_3}). For an instantaneous
perturbation, a related to interaction with the boundary effect has been
observed in lattice dynamics~\cite{RMJ, gavrilov2023example} and referred to as
anti-localization~\cite{gavrilov2023example}; the particle
velocity at the boundary decays as~$t^{-3/2}$. Squaring
this result suggests a~$t^{-3}$ decay of the corresponding
mean-square velocity, associated with the kinetic temperature.} The farther the source is from the boundary, the weaker the interference of the reflected and incident waves becomes, because their energy is lost due to dispersion and~(a fortiori) viscosity.
\par We introduce a function~$g(t,x,\xi)$ such that
\begin{equation}\label{ContPH}
    ag(t,na,ja)\defeq \dot{\mathit{\Phi}}_{n,j}^2.
\end{equation}
Having replaced~$\sum_j^\infty$ by~$\frac{1}{a}\int_0^\infty ... \mathrm{d}\xi$ and taken into account both the fast and slow processes and also~(\ref{ContSource}), we write the \edited{following expression} for the kinetic temperature as follows:
\begin{equation}\label{CONTSOL}
\begin{split}
   \DS T(t,x)=2\int_0^\infty\int_{\frac{|x-\xi|}{v_s}}^t \chi(\tau,\xi) e^{-2\eta(t-\tau)} G(t-\tau,|x-\xi|) \mathrm{d}\tau\mathrm{d}\xi\\[3mm]+2\int_0^\infty \int_{\frac{x+\xi+a}{v_s}}^t \chi(\tau,\xi) e^{-2\eta(t-\tau)} G(t-\tau,x+\xi+a) \mathrm{d}\tau\mathrm{d}\xi
   +\Delta T_\mathrm{B},
\end{split}
\end{equation}
\begin{equation}\label{BOUNDARY_LAYER_ADD_2}
    \Delta T_\mathrm{B}(t,x)=2\int_0^\infty \int_{\frac{x+\xi+a}{v_s}}^t \chi(\tau,\xi)e^{-2\eta(t-\tau)}\Delta g(t-\tau,x,\xi)\vert_{\mathrm{q}=1}\mathrm{d}\tau\mathrm{d}\xi,
\end{equation}
\begin{equation}\label{BOUNDARY_LAYER_INTEGRAND}
 \Delta g(t,x,\xi)\defeq \frac{\mathrm{q}\sin \left(\left(\mathcal{W}\left(|x-\xi|\right)+\mathcal{W}\left(x+\xi+a\right)\right)t\right)+\cos\left(\left(\mathcal{W}\left(|x-\xi|\right)-\mathcal{W}\left(x+\xi+a\right)\right)t\right)}{\pi\sqrt[4]{\left(v_s^2t^2-(x-\xi)^2\right)\left(v_s^2t^2-(x+\xi+a)^2\right)}}.
\end{equation}
Here~$\Delta T_\mathrm{B}$ is a term corresponding to the influence of the free boundary on thermal processes; $\mathrm{q}$ is a constant, which is equal to~$1$ if the fast oscillation term is included in the expression for the~$\Delta T_\mathrm{B}$ and~$0$ otherwise;~$\mathcal{W}(t,x)$ is a continuous function of the characteristic frequency such as 
\begin{equation}\label{PSI_C}
    \mathcal{W}(t,na)\equiv \mathcal{W}_{n}(t).
\end{equation} 
According to the analysis of the sudden point heat supply~(see Sect.~\ref{sect_4_4}), neglecting the fast term in~(\ref{BOUNDARY_LAYER_INTEGRAND}), i.e., putting~$\mathrm{q}=0$, results in a loss in the accuracy of the solution for the kinetic temperature at and near the boundary. On the other hand, in general, at~$\mathrm{q}=1$ the solution for kinetic temperature \edited{oscillates between lattice sites}. Therefore, we classify the solution for kinetic temperature~(\ref{CONTSOL}) as a \textit{discrete-continuum} \edited{solution}. The steady-state nonequilibrium solution for kinetic temperature, if it exists, is defined as~$T^\infty(x)\defeq T(\infty,x)$. Note that the discrete-continuum solution for the kinetic temperature in the form is valid only for weakly dissipative case~($\eta\ll \omega_e$), while the symmetrical continuum solution is valid for arbitrary~$\eta$.  
\subsection{General case. Example:~sudden point heat supply}\label{sect_4_4}

\par Consider the sudden point heat supply with the constant intensity:
\begin{equation}\label{DIRAC_SUPPLY}
    \chi(t,x)=\chi^0a\delta(x-h)H(t),\quad h=ja,
\end{equation}
where~$\chi^0$ is a constant of~$\mathrm{K}/\mathrm{s}$ dimension. The corresponding expression for the source~$\chi$ in the discrete form is written as~
\begin{equation}\label{SUPPLY_KRONECK}
    \chi_n(t)=\chi^0\delta_{n,j}H(t).
\end{equation}
The discrete solution for the kinetic temperature is obtained by substitution of~Eq.~(\ref{SUPPLY_KRONECK}) into~Eq.~(\ref{KinTT2}) with taking into account~(\ref{SOL_EQ26}), which yields 
\begin{equation}\label{DISCRETE_SOL_POINT_SUPP}
    T_n(t)=2\chi^0\int_0^t e^{-2\eta \tau}\left(J_{2|n-j|}(2\omega_e\tau)+J_{2(n+j+1)}(2\omega_e\tau)\right)^2\mathrm{d}\tau.
\end{equation}
In the limit~$t\rightarrow \infty$, the equation for~$T_n$ may be obtained in terms of the hypergeometric functions~\cite{Masirevic}, in a closed but rather cumbersome form.
\par The symmetrical continuum solution is obtained by substitution of Eq.~(\ref{DIRAC_SUPPLY}) into Eq.~(\ref{TEMP_CLASSIC_CONT}) yields 
\begin{equation}\label{CLASS_CONT_SOL_POINT_SUPP}
\begin{split}T(t,x)=\frac{\chi^0aH(v_st-|x-h|)}{\pi}\int_{\frac{|x-h|}{v_s}}^t\frac{e^{-2\eta \tau}\mathrm{d}\tau}{\sqrt{v_s^2\tau^2-(x-h)^2}}\\+\frac{\chi^0aH(v_st-|x+h|)}{\pi}\int_{\frac{x+h}{v_s}}^t\frac{e^{-2\eta \tau}\mathrm{d}\tau}{\sqrt{v_s^2\tau^2-(x+h)^2}}.
\end{split}
\end{equation}
When viscosity is present, a nonequilibrium steady state exists. The symmetrical continuum solution has the following form:
\begin{equation}\label{CLASS_STAT_CONT}
    T^\infty(x)=\frac{\chi^0}{\pi\omega_e}\left(K_0\left(\frac{2\eta|x-h|}{v_s}\right)+K_0\left(\frac{2\eta(x+h)}{v_s}\right)\right).
\end{equation}
\par The discrete-continuum solution for the kinetic temperature is obtained by substitution of Eq.~(\ref{DIRAC_SUPPLY}) into Eq.~(\ref{CONTSOL}), which yields
\begin{equation}\label{DISCR_CONT_SOL_POINT_SUPP}
\begin{split}
T(t,x)=\frac{\chi^0aH(v_st-|x-h|)}{\pi}\int_{\frac{|x-h|}{v_s}}^t\frac{e^{-2\eta \tau}\mathrm{d}\tau}{\sqrt{v_s^2\tau^2-(x-h)^2}}\\+\frac{\chi^0aH(v_st-|x+h+a|)}{\pi}\int_{\frac{x+h+a}{v_s}}^t\frac{e^{-2\eta \tau}\mathrm{d}\tau}{\sqrt{v_s^2\tau^2-(x+h+a)^2}}+\Delta T_\mathrm{B},  
\end{split}
\end{equation}
\begin{equation}\label{DELTA_R_B_NEU}
    \Delta T_\mathrm{B}(t,x)=2\chi^0aH(v_st-|x+h+a|)\int_{\frac{x+h+a}{v_s}}^t e^{-2\eta \tau} \Delta g(\tau,x,h)|_{\mathrm{q}=1}\mathrm{d}\tau.
\end{equation}
Therefore, the stationary solution,~$T^\infty$, is written as
\begin{equation}\label{DISCR_STAT_CONT}
\begin{split} T^\infty(x)=\frac{\chi^0}{\pi\omega_e}\left(K_0\left(\frac{2\eta|x-h|}{v_s}\right)+K_0\left(\frac{2\eta(x+h+a)}{v_s}\right)\right)\\[1mm]+2\chi^0a\int_{\frac{x+h+a}{v_s}}^\infty e^{-2\eta \tau} \Delta g(\tau,x,h)|_{\mathrm{q}=1}\mathrm{d}\tau.
\end{split}
\end{equation}
The principal difference in the steady state solution for the kinetic temperature between the discrete-continuum and symmetrical continuum descriptions is that the discrete-continuum solution takes into account the interactions of the thermal wave with the boundary~\edited{(namely, interference of the incident and reflected thermal waves)}, which the symmetrical continuum solution does not~\edited{retain}. 
\par Now we find an approximate solution for the far field~(for~$x\gg h+a$) of kinetic temperature. Far from the source, the symmetrical continuum steady-state solution~(Eq.~(\ref{CLASS_STAT_CONT})) may be estimated by using the following expansion of the modified Bessel function of the second kind at large argument~(see~Eq.~$9.7.2$ in~\cite{Abramowitz}):

\begin{equation}\label{MOD_BESS_EXP}
K_0(\lambda)=e^{-\lambda}\sqrt{\frac{\pi}{2\lambda}}\left(1+O\left(\lambda^{-1}\right)\right),\qquad \lambda\rightarrow \infty.
\end{equation}
Substitution of the first term in~(\ref{MOD_BESS_EXP}) into~(\ref{CLASS_STAT_CONT}) yields
\begin{equation}\label{FAR_KIN_CLASS}
    T^\infty(x)\sim \frac{\chi^0}{2\omega_e}\left(\sqrt{\frac{v_s}{\pi \eta(x-h)}}e^{-\frac{2\eta(x-h)}{v_s}}+\sqrt{\frac{v_s}{\pi \eta(x+h)}}e^{-\frac{2\eta(x+h)}{v_s}}\right),\quad x\rightarrow \infty.
\end{equation}
\par Approximation of the far-field discrete-continuum solution implies asymptotic estimate of the last term in~(\ref{DISCR_STAT_CONT}), $\Delta T_\mathrm{B}$. Instead of using well-known asymptotic techniques~\cite{Temme} we estimate~(\ref{DISCR_STAT_CONT}) in easier way, knowing the behavior of the function~$\Delta g$. If the source is not at the boundary, then the~$\Delta g$ for~$x\gg h+a$ represents an integral \edited{of a rapidly} oscillating function, because the characteristic frequencies of the incident and reflected thermal waves are not close to each other~(see Sect.~\ref{S4_2}). Therefore,~$\Delta g$ may be neglected, and the expression for the far-field kinetic temperature is written as
\begin{equation}\label{FAR_KIN_DC}
    T^\infty(x)\vert_{h\neq 0}\approx \frac{\chi^0}{2\omega_e}\left(\sqrt{\frac{v_s}{\pi \eta(x-h)}}e^{-\frac{2\eta(x-h)}{v_s}}+\sqrt{\frac{v_s}{\pi \eta(x+h+a)}}e^{-\frac{2\eta(x+h+a)}{v_s}}\right).
\end{equation}
Thus, for~$h\neq 0$ the far-field discrete-continuum solution \edited{does not differ significantly} from the far-field symmetrical continuum solution~(\ref{FAR_KIN_CLASS}). For~$h=0$, the estimation~(\ref{FAR_KIN_DC}) is not valid because the last term in~(\ref{DISCR_STAT_CONT}) may not be negligible. However, we neglect fast-oscillating terms in~$\Delta g$, substituting~$\mathrm{q}=0$. Simplifying~Eq.~(\ref{BOUNDARY_LAYER_INTEGRAND}) by using~Eq.~(\ref{DELTA_FREQ}), we write the following expression for~$\Delta T_\mathrm{B}$:
\begin{equation}\label{DELTA T_B_2}
     \Delta T_\mathrm{B}\approx\frac{2\chi^0a}{\pi}\int_{\frac{x+a}{v_s}}^\infty \frac{e^{-2\eta \tau}\left(-1+\frac{(2x+a)^2}{2v_s^2\tau^2}\right)\mathrm{d}\tau}{\sqrt[4]{(v_s^2\tau^2-x^2)(v_s^2\tau^2-(x+a)^2)}}.
\end{equation}
\edited{Introducing}~$\Tilde{\tau}=\tau-\frac{x+a}{v_s}$, Eq.~(\ref{DELTA T_B_2}) is rewritten as
\begin{equation}\label{DELTA_T_SUPP_2}
    \Delta T_\mathrm{B}\approx\frac{2\chi^0a}{\pi}e^{-\frac{2\eta(x+a)}{v_s}}\bigint_0^\infty \frac{e^{-2\eta\Tilde{\tau}}\left(-1+\frac{(2x+a)^2}{2(v_s\Tilde{\tau}+x+a)^2}\right)\mathrm{d}\Tilde{\tau}}{\sqrt[4]{\left(\left(v_s\Tilde{\tau}+x+a\right)^2-x^2\right)\left(\left(v_s\Tilde{\tau}+x+a\right)^2-(x+a)^2\right)}}.
\end{equation}
Expanding the integrand in~(\ref{DELTA_T_SUPP_2}) at~$x=\infty$ neglecting the terms of order of~$x^{-\frac{3}{2}}$ and higher, and, using the identity 
\begin{equation}\label{IDENT_INTEGR}
    \int_0^\infty\frac{e^{-2\eta \Tilde{\tau}}\mathrm{d}\Tilde{\tau}}{\sqrt[4]{v_s\Tilde{\tau}(v_s\Tilde{\tau}+a)}}=\sqrt[4]{\frac{2a}{v_s^3\eta}}\frac{e^{\frac{\eta}{\omega_e}}\sqrt{\pi}K_{\frac{1}{4}}\left(\frac{\eta}{\omega_e}\right)}{\Gamma\left(\frac{1}{4}\right)},
\end{equation}
where~$\Gamma(y)\defeq \int_0^\infty e^{-t}t^{y-1}\mathrm{d}t$ is the Gamma function, we have 
\begin{equation}\label{DELTA_T_SUPP_3}
    \Delta T_\mathrm{B}\simeq\frac{\sqrt[4]{8}\chi^0a}{\sqrt{\pi}\Gamma\left(\frac{1}{4}\right)}\sqrt[4]{\frac{a}{v_s^3\eta}}K_{\frac{1}{4}}\left(\frac{\eta}{\omega_e}\right)e^{-\frac{\eta}{\omega_e}}\frac{e^{-\frac{2\eta x}{v_s}}}{\sqrt{x}},\quad x\rightarrow\infty.
\end{equation}
Therefore, the expression for the far-field kinetic temperature at the source at the boundary has the form 
\begin{equation}\label{FAR_KIN_T_2}
\small
    T^\infty(x)\vert_{h= 0}\approx \frac{\chi^0}{2\omega_e}\left(\left(\sqrt{\frac{v_s}{ \eta }}+\frac{2}{\Gamma\left(\frac{1}{4}\right)}\sqrt[4]{\frac{8v_sa}{\eta}}K_{\frac{1}{4}}\left(\frac{\eta}{\omega_e}\right)e^{-\frac{\eta}{\omega_e}}\right)\frac{e^{-\frac{2\eta x}{v_s}}}{\sqrt{\pi x}}+\sqrt{\frac{v_s}{\pi \eta(x+a)}}e^{-\frac{2\eta(x+a)}{v_s}}\right).
\end{equation}
Thus, we have obtained an analytical solution for the kinetic temperature in the continuum limit. \edited{Simplifications were obtained only in special cases or for the far-field stationary profile} of the kinetic temperature. 
\par We compare analytical results for the kinetic temperature, corresponding to the discrete, \edited{symmetrical continuum and discrete-continuum solutions} with the corresponding numerical solutions. \edited{The numerical simulations are performed for a finite chain with
\(N=512\) particles, time step \(\omega_e\Delta t=0.01\), excitation
amplitude~\(b/(v_s\sqrt{\omega_e})=1\), and \(N_\mathrm{R}=10^5\) independent realizations, by averaging over which the mathematical expectation in Eq.~(\ref{eq4}) is replaced~(see the Appendix~\ref{CCC} for details}. The time intervals and
observation sites shown below are chosen so that waves reflected from
the remote end do not affect the displayed results.  The integrals \edited{appearing in the analytical solution} are calculated using the trapezoidal rule. In particular, the steady-state solution for the dissipative case is obtained as the non-stationary one for a sufficiently long time after the beginning of the supply~(we take~$\omega_et=500$).  
Here, we investigate the obtained results, corresponding to both interior source~(for instance, $j=5$) and one at the boundary. 

\subsubsection{The non-dissipative case}\label{sect_4_4_1}
\par We consider the non-dissipative case~($\eta\rightarrow0+$). Then, the symmetrical continuum solution for the kinetic temperature has the form: 
\begin{equation}\label{NOVISC_CLASS_CONT_SOL}
\begin{split}
     T(t,x)=\frac{\chi^0H(v_st-|x-\xi|)}{\pi\omega_e}\ln{\frac{v_st+\sqrt{v_s^2t^2-(x-\xi)^2}}{|x-\xi|}}\\+\frac{\chi^0H(v_st-(x+\xi))}{\pi\omega_e}\ln{\frac{v_st+\sqrt{v_s^2t^2-(x+\xi)^2}}{x+\xi}},
\end{split}
\end{equation}
from which logarithmic growth of~$T(t,x)$ follows. In turn, the discrete-continuum solution is obtained by substitution of~$\eta=0$ into Eq.~(\ref{DISCR_CONT_SOL_POINT_SUPP}), \edited{which} yields coincidence of the first two terms, denoting the contributions from incident and reflected waves with Eq.~(\ref{NOVISC_CLASS_CONT_SOL}), where~$x+\xi$ is replaced by~$x+\xi+a$. The discrete solution is obtained by substitution of~$\eta=0$ into Eq.~(\ref{DISCRETE_SOL_POINT_SUPP}).

\begin{figure}[htb]
\begin{minipage}[htb]{0.5\linewidth}
\center{\includegraphics[width=0.95\linewidth]{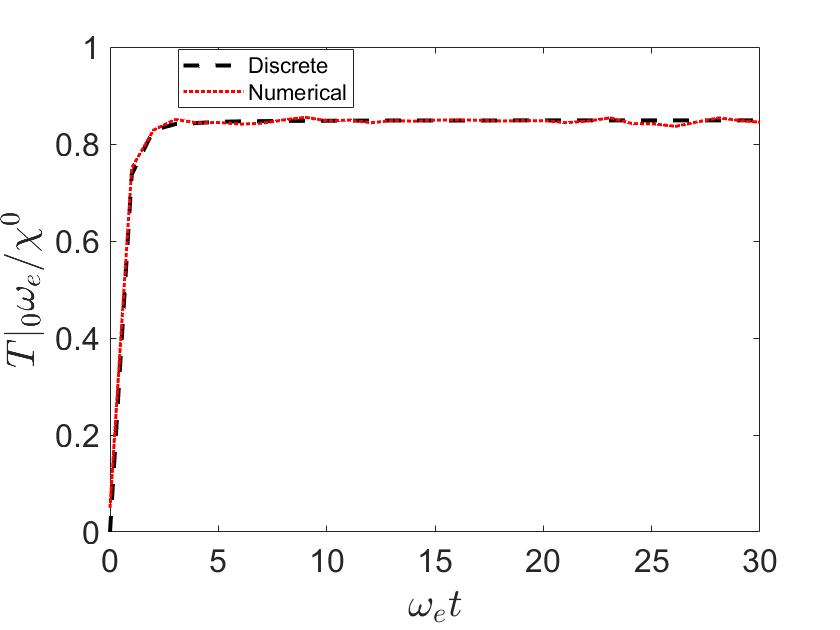}}\\A
\end{minipage}
\begin{minipage}[htb]{0.5\linewidth}
\center{\includegraphics[width=0.95\linewidth]{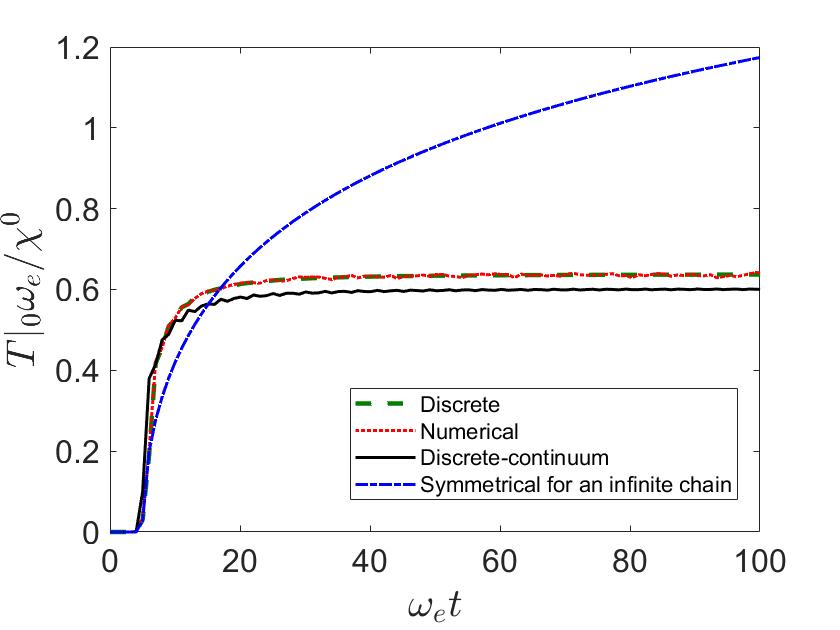} }\\B
\end{minipage}
\caption{Evolution of the kinetic temperature at the boundary~($n=0$) in the \edited{non-dissipative~($\eta = 0$)} semi-infinite chain subjected to~\edited{sudden constant-intensity heat supply.} \edited{A: Source at the free end,
\(j=0\)}. \edited{B: Interior source, \(j=5\).} \edited{The discrete solution~((\ref{DISCRETE_SOL_POINT_SUPP}), dashed line), discrete-continuum solution~(Eqs.~(\ref{CONTSOL}),~ (\ref{BOUNDARY_LAYER_ADD_2}), solid line), symmetrical continuum solution~(Eq.~(\ref{NOVISC_CLASS_CONT_SOL}), dash-dotted line) and numerical solution~(dotted line)} are demonstrated.}
\label{fig9}
\end{figure}

We study the evolution in time of the kinetic temperature and the influence of the free boundary on the latter. Consider the kinetic temperature at the boundary. The comparison of analytical and numerical solutions for the latter is shown in Fig.~\ref{fig9}.

\begin{figure}[htb]
\center{\includegraphics[width=0.65\linewidth]{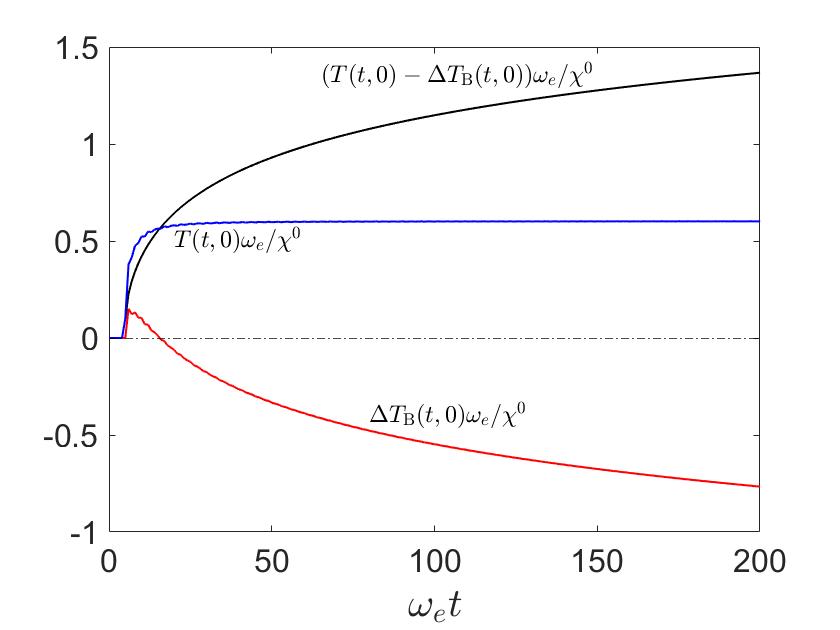}}
\caption{\edited{Decomposition of the discrete-continuum solution kinetic temperature at the
free end~(\(n=0\)) in the non-dissipative~(\(\eta=0\)) semi-infinite chain for a sudden constant-intensity interior source
at \(j=5\). Discrete-continuum solution given by
Eqs.~(\ref{CONTSOL})---(\ref{BOUNDARY_LAYER_INTEGRAND})~(blue line); the free boundary contribution $\Delta T_\mathrm{B}$~(Eq.~(\ref{BOUNDARY_LAYER_ADD_2}), red line) and the sum of the contributions from the incident and reflected waves~(Eq.~(\ref{CONTSOL}), black line)} are shown.}
\label{fig10}
\end{figure}

\edited{It is seen} that, at short times, the discrete and discrete-continuum solutions for the kinetic temperature at the boundary grow faster than the symmetrical continuum one~(see Fig.~\ref{fig9}B). However, at large times, growth of the discrete solution \edited{slows so markedly that the temperature appears to approach an apparent local plateau.} To explain the reason for this behavior of these solutions and determine whether the corresponding process is a transition to the local stationary state or an anomalously slow growth of the kinetic temperature, we investigate the discrete-continuum solution, which behaves \edited{very similarly} to the discrete one, although their \edited{large-time values} differ~(see Fig.~\ref{fig9}B). We analyze the dependence of the first two terms in Eq.~(\ref{DISCR_CONT_SOL_POINT_SUPP}), which is shown in Fig.~\ref{fig10}.
We see that, firstly, after the incident thermal wave reaches the boundary, the discrete-continuum solution for the kinetic temperature grows faster than the sum of its contribution from the incident and reflected waves due to~$\Delta T_\mathrm{B}>0$. \edited{Secondly, the growth rate of the kinetic temperature decreases after
\(\Delta T_{\mathrm{B}}\) changes sign.} At large times, the absolute value of~$\Delta T_\mathrm{B}$ grows in time similarly to the sum of the aforementioned contributions. \edited{The time derivatives of this sum and $\Delta T_\mathrm{B}$ have
opposite signs and comparable magnitudes over the time interval shown.
Their partial compensation produces the pronounced slowing of the
kinetic-temperature growth. This numerical observation is consistent
with an \edited{apparent local plateau}, but does not by itself establish the
large-time limit. The present computations further indicate that the boundary
temperature may exhibit such slowing irrespective of whether the
source is located at the free end or in the interior. For an interior
source, a similar trend is observed at sites between the source and
the boundary.}

\par
\begin{figure}[htb]
\begin{minipage}[htb]{0.5\linewidth}
\center{\includegraphics[width=0.95\linewidth]{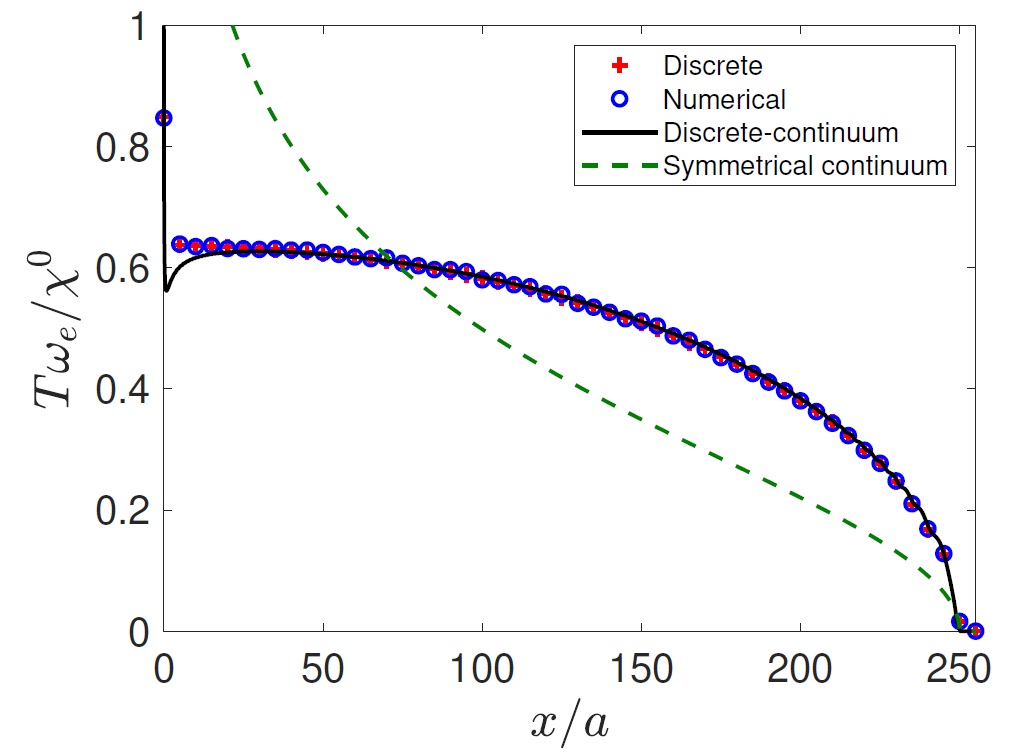}}\\A
\end{minipage}
\begin{minipage}[htb]{0.5\linewidth}
\center{\includegraphics[width=0.95\linewidth]{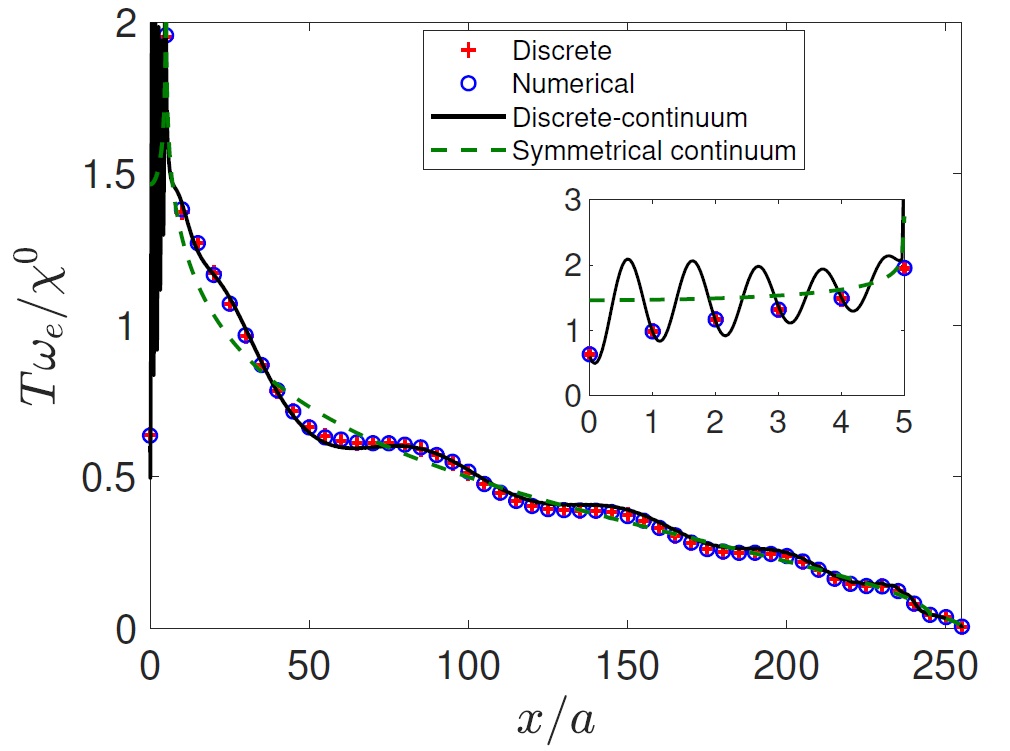} }\\B
\end{minipage}
\caption{\edited{Spatial profiles} of the kinetic temperature in the \edited{non-dissipative~($\eta=0$)} semi-infinite chain subjected to~\edited{sudden constant-intensity heat supply at~$\omega_et=250$.} \edited{A: Source at the free end,
\(j=0\)}. \edited{B: Interior source, \(j=5\).} \edited{The discrete solution~(Eq.~(\ref{DISCRETE_SOL_POINT_SUPP}), <<crosses>>), discrete-continuum solution~((Eqs.~(\ref{CONTSOL}),~ (\ref{BOUNDARY_LAYER_ADD_2}), solid line), symmetrical continuum solution~(Eq.~(\ref{NOVISC_CLASS_CONT_SOL}), dashed line) and numerical solution~(<<circles>>)} are demonstrated.}
\label{fig11}
\end{figure}

\begin{figure}[htb]
\begin{minipage}[htb]{0.5\linewidth}
\center{\includegraphics[width=0.95\linewidth]{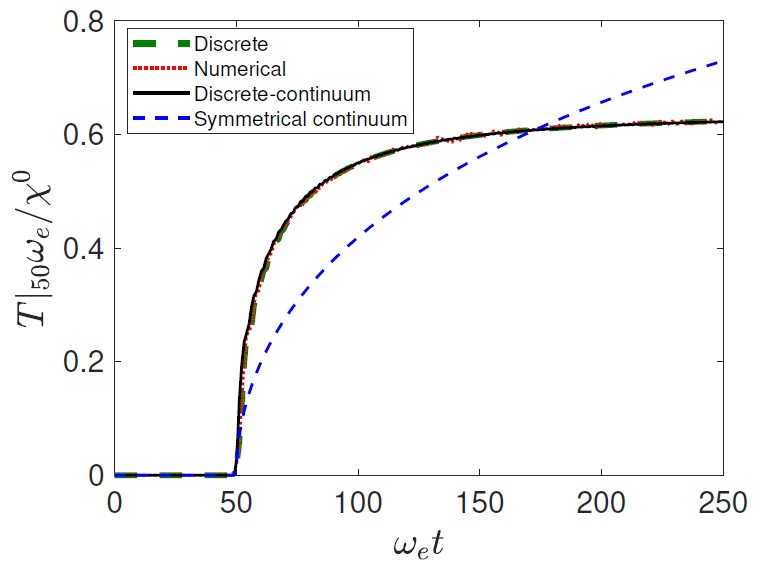} }\\A
\end{minipage}
\begin{minipage}[htb]{0.5\linewidth}
\center{\includegraphics[width=0.95\linewidth]{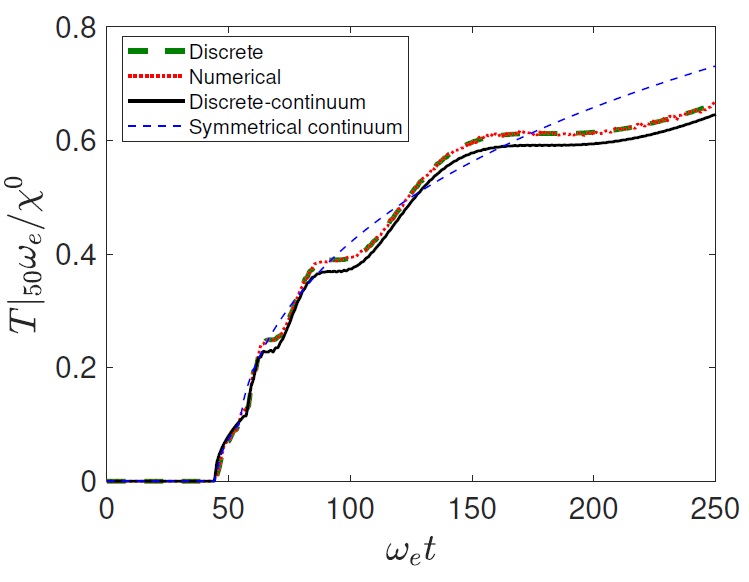} }\\B
\end{minipage}
\caption{Evolution of the kinetic temperature at the $50$-th point in the \edited{non-dissipative~($\eta = 0$)} semi-infinite chain subjected to~\edited{sudden constant-intensity heat supply.} \edited{A: Source at the free end,
\(j=0\)}. \edited{B: Interior source, \(j=5\).} \edited{The discrete solution~((\ref{DISCRETE_SOL_POINT_SUPP}), bold dashed line), discrete-continuum solution~(Eqs.~(\ref{CONTSOL}),~ (\ref{BOUNDARY_LAYER_ADD_2}), solid line), symmetrical continuum solution~(Eq.~(\ref{NOVISC_CLASS_CONT_SOL}), dashed line) and numerical solution~(dotted line)} are demonstrated.}
\label{fig12}
\end{figure}

 \edited{The observed compensation suggests that the boundary term is essential
for the finite-time behavior near the free end. A rigorous asymptotic
analysis is nevertheless required to determine whether the
corresponding discrete temperature has a finite limit as
\(t\to\infty\). This analysis remains beyond the scope of \edited{the} present paper.} The profiles of the kinetic temperature at fixed moment of time are shown\footnote{To avoid a jumble of symbols in Fig.\ref{fig11}, we plot the discrete and numerical solutions not at every point, but rather at every fifth.} in Fig.~\ref{fig11}.
It is seen in Fig.~\ref{fig11} that, when the supply occurs at the point~$j=0$, the kinetic temperature at the points far from the boundary grows sufficiently more slowly, than in the case of supply at the point~$j=5$. \edited{To examine this difference}, we investigate the evolution of the kinetic temperature at an arbitrary point far from the boundary. Comparison of the corresponding solutions is shown in Fig.~\ref{fig12}.
It is seen that, in the case of the supply at the point~$j=0$, the kinetic temperature \edited{exhibits pronounced
large-time slowing and appears to approach a local plateau}~(see Fig.~\ref{fig12}A), in contrast to the kinetic temperature in the case of supply outside the boundary~(see Fig.~\ref{fig12}B). Having analyzed~(in the similar way) behavior of the kinetic temperature at several points, we conclude that, \edited{ when the source is located at the free
end, the discrete solution exhibits a gradual large-time slowdown
throughout the heated region. Over the time interval considered, this
behavior is consistent with the progressive formation of a local
nonequilibrium plateau.}
 From Figs.~\ref{fig9},~\ref{fig11},~\ref{fig12} it is seen~\edited{that the discrete-continuum solution captures
important finite-time boundary effects that are absent in the
symmetrical continuum solution. However, for an interior source its
large-time behavior does not quantitatively reproduce the discrete
solution throughout the near-boundary region. The symmetrical
continuum solution is therefore not suitable for describing this
boundary-sensitive regime.}

\subsubsection{The \edited{weakly} dissipative case}\label{sect_4_4_2}

\begin{figure}[htb]
\center{\includegraphics[width=0.65\linewidth]{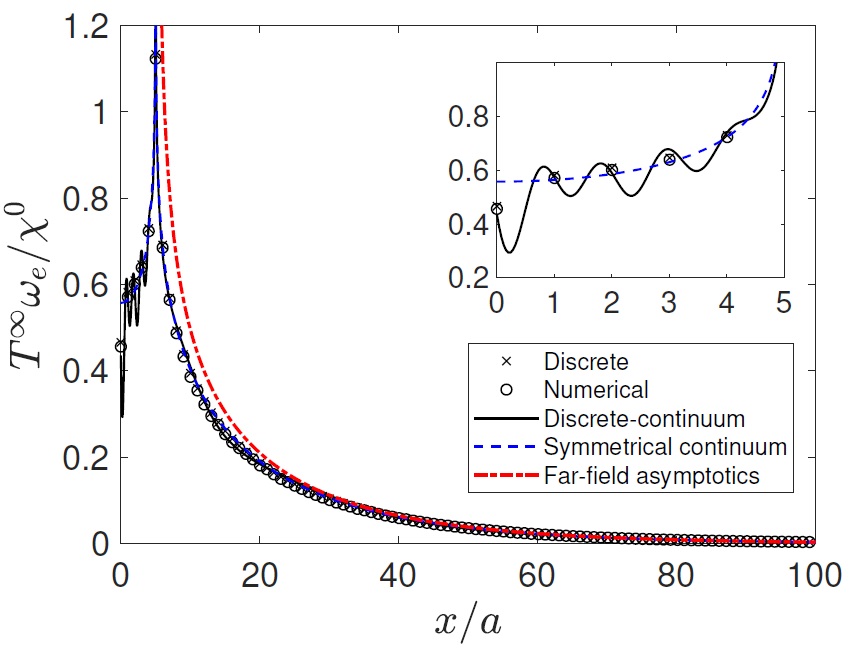}}
\caption{\edited{Large-time kinetic-temperature profile for a sudden constant-intensity interior source, \(j=5\), in the weakly dissipative case
~(\(\eta/\omega_e=0.02\)). The discrete solution~(Eq.~(\ref{DISCRETE_SOL_POINT_SUPP}), <<crosses>>), discrete-continuum solution~((Eqs.~(\ref{CONTSOL}),~ (\ref{BOUNDARY_LAYER_ADD_2})), solid line), symmetrical continuum solution~(Eq.~(\ref{CLASS_CONT_SOL_POINT_SUPP}), dashed line), far-field asymptotics~(Eq.~(\ref{FAR_KIN_DC}), dash-dotted line) and numerical solution~(<<circles>>)} are demonstrated.}
\label{stat_h5}
\end{figure}

\begin{figure}[htb]
\center{\includegraphics[width=0.65\linewidth]{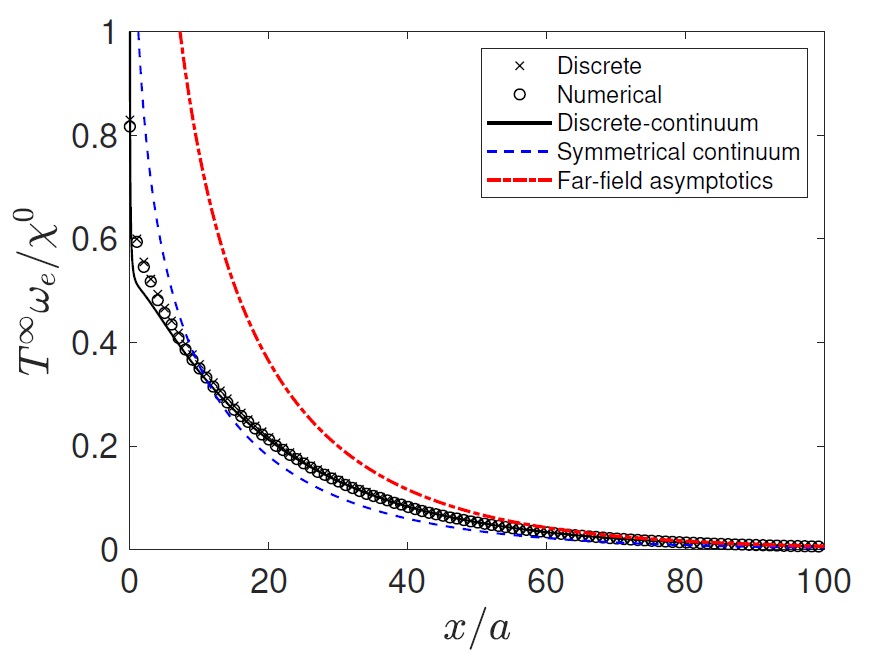}}
\caption{\edited{Large-time kinetic-temperature profile for a sudden constant-intensity source at the free end, \(j=0\), in the weakly dissipative case
~(\(\eta/\omega_e=0.02\)). The discrete solution~(Eq.~(\ref{DISCRETE_SOL_POINT_SUPP}), <<crosses>>), discrete-continuum solution~((Eqs.~(\ref{CONTSOL}),~ (\ref{BOUNDARY_LAYER_ADD_2})), solid line), symmetrical continuum solution~(Eq.~(\ref{CLASS_CONT_SOL_POINT_SUPP}), dashed line), far-field asymptotics~(Eq.~(\ref{FAR_KIN_T_2}), dash-dotted line) and numerical solution~(<<circles>>)} are demonstrated.}
\label{stat_h0}
\end{figure}

In the presence of weak dissipation, the transition to the nonequilibrium steady state is expected at all points. The kinetic temperature as a function of time tends to some stationary value~(according to Eq.~(\ref{DISCRETE_SOL_POINT_SUPP})). For large distances from the source or long times, the kinetic temperature approaches a steady-state profile.
The solutions for the kinetic temperature, corresponding to the case of the supply at the point~$j=5$ are shown in Fig.~\ref{stat_h5}.
We see that, firstly, far from the boundary, the discrete-continuum and symmetrical continuum solutions for the kinetic temperature are in good agreement. However, near the boundary, the discrete-continuum solution describes the kinetic temperature field more accurately than the \edited{symmetrical} continuum solution, although the difference between the solutions is not \edited{significant}. The far-field asymptotic solution~(\ref{FAR_KIN_DC}) describes the kinetic temperature field~(both the discrete-continuum and continuum solutions) rather accurately for~$x/a>30$. 
The other situation is when the supply is at the boundary~(see Fig.~\ref{stat_h0}).
We see that the behavior of the symmetrical continuum solution for the stationary field of the kinetic temperature is even qualitatively different from the behavior of the discrete-continuum solution. In contrast to the case of \edited{an interior source}, the qualitative difference of the solution is observed not only near the boundary but everywhere. The far-field asymptotic solution~(\ref{FAR_KIN_T_2}) describes the kinetic temperature field rather accurately for~$x/a>60$. One can observe in Fig.~\ref{stat_h0}, that, near the boundary, the discrete and discrete-continuum solutions change more abruptly than the symmetrical continuum solution. A convincing explanation for the latter observation remains an open problem.

\section{Conclusions}\label{sect_6}

Heat transport in a semi-infinite free end Hooke chain immersed in a weakly viscous environment and having an arbitrary heat source has been investigated. We started with Eqs.~(\ref{eq1}) for the stochastic dynamics of the chain and  derived an exact expression for the kinetic temperature~(Eq.~(\ref{KinTT})) along with its simplified weakly dissipative version~(Eq.~(\ref{KinTT2})), termed the discrete solution. We continualized this using two methods. The first applies the symmetry principle for continuum solutions from~\cite{Sinf2023}, yielding the symmetrical continuum solution.  The second, based on a method proposed in~\cite{Gavr2022} for discrete problems, estimates the fundamental thermal solution at the incident and reflected wave fronts as~$t \rightarrow \infty$, producing contributions from incident and reflected waves, and a cross term, which is responsible for the interactions of these waves. For sudden point heat supply, \edited{the exact discrete solution reveals a boundary-induced large-time slowdown even in the non-dissipative case \(\eta=0\).  The symmetrical continuum solution misses this effect and instead predicts unbounded logarithmic growth. The discrete-continuum representation retains a term, related to interactions with the boundary and captures important finite-time deviations from the symmetrical continuum result. However, for an interior source it does not quantitatively reproduce the apparent large-time plateau of the discrete solution near the boundary. Establishing the precise large-time asymptotics of the discrete solution, and deriving a uniform continuum approximation in that regime, remain open problems.}

\edited{Both continuum representations have limitations. The symmetrical
continuum solution and the
discrete-continuum solution are singular at the supply point.} We lack a clear method to resolve this; replacing the Dirac delta source~(\ref{DIRAC_SUPPLY}) with a Gaussian one did not yield convincing results. We suggest that the singularities seen here and in~\cite{Gavr2018,Gavr2020,Gavr2021} could be addressed using nonlocal function analysis from~\cite{VASLUR}. The second drawback is the \edited{oscillatory} behavior of the discrete-continuum solution between the particle points~(especially as it manifests near the boundary during the constant intensity heat supply). Although this drawback does not affect the nature of heat propagation itself, it can manifest when the discrete-continuum solution is used as a constitutive relation in thermoelastic  problems~(see~\cite{Kuz20202, Trunova}).

\edited{A natural extension of the current work is to replace the white-noise driving by
Ornstein--Uhlenbeck colored noise, thereby introducing a finite
correlation time~\footnote{This idea was proposed by M.V. Golub in private communication.}. Other extensions are as follows.}  The first is to solve the analogous problem for the heat flux. The second is to generalize the approaches proposed here to higher-dimensional lattices. In particular, developing a theory of (quasi-)ballistic heat transport in higher-dimensional lattices may provide a clearer description of experiments such as those based on transient thermal grating~\cite{huberman2019, ZH} or phonon focusing observations \cite{Northrop, Northrop2, Koos} than \edited{conventional} continuum models. The third is to incorporate retardation in interactions to achieve a clearer transition from mechanics to thermodynamics \cite{ZAKH}. \edited{The first-order retardation} effects could be explored by replacing spring elements with spring-damper ones in the chain. 
\par \edited{An important extension is to introduce mechanisms that convert the
ballistic transport of the harmonic chain into Fourier-like transport.
Such mechanisms may include anharmonicity~\cite{Kuz2020}, bond breaks~\cite{Kuz2024} or
energy-conserving stochastic scattering~\cite{Palla2020}. Deriving a continuum limit for the present
semi-infinite geometry with such a mechanism, and identifying the
conditions for a Fourier-type macroscopic limit, is also an important
open problem.}

\textbf{Acknowledgments}
The author is grateful to V.A. Kuzkin, A.M. Krivtsov, S.N. Gavrilov, E.V. Shishkina, A.A. Sokolov, E.F. Grekova, Jie Chen, \edited{M.V. Golub}, \edited{Yu. V. Petrov} and M.M. Khasanov for valuable discussions. \edited{Comments of the referees are highly appreciated}. The results of the project~<<FR-2025-75>>, carried out within the framework of the Basic Research Program at HSE University, are presented in this work.
\begin{appendices}

\section{Derivation of Eq.~(\ref{eq12})}\label{AAA}

Make the following change of variable in Eq.~(\ref{eq10}):
\begin{equation}\label{A_EQ_1}
    \Vect{y}=e^{-\Vect{A}t}\hat{\Vect{u}}.
\end{equation}
Then, to find $\mathrm{d}\Vect{y}$, we use the It{\^o} lemma~(see Eq.~(6.12) in~\cite{Stepanov}). For the~$s$-th  component of~$\mathrm{d}\Vect{y}$ it can be written as
\begin{equation}\label{A_EQ_2}
    \mathrm{d}y_s=\Bigg(\frac{\partial y_s}{\partial t}+\frac{\partial y_s}{\partial {\hat{u}}_\alpha}(\Vect{A} \hat{\Vect{u}})_\alpha+\frac{1}{2}\frac{\partial^2 y_s }{\partial{\hat{u}}_\alpha \partial{\hat{u}}_\beta }B_{\alpha \gamma}B_{\beta \gamma}\Bigg) \mathrm{d}t+\frac{\partial y_s}{\partial{\hat{u}}_\alpha}B_{\alpha \gamma}\mathrm{d}W_\gamma,
\end{equation}
where summation over repeating indices is implied~$\left(\sum_{\alpha}X_\gamma Y_\gamma\right)$. Substituting of each component of~$\Vect{y}$ from Eq.~(\ref{A_EQ_1}) to Eq.~(\ref{A_EQ_2}) one concludes the first term of Eq.~(\ref{A_EQ_2}) to be equal to zero~$\forall s$. Therefore, we have the following equation:
\begin{equation}\label{A_EQ_3}
    \mathrm{d}\Vect{y}=e^{-\Vect{A}t}\Vect{B}\mathrm{d}\Vect{W},
\end{equation}
whereas, due to the initial condition~(\ref{eq11})
\begin{equation}\label{A_EQ_4}
    \Vect{y}=\int_0^t e^{-\Vect{A}\tau}\Vect{B}(\tau)\mathrm{d}\Vect{W}.
\end{equation}
The reverse substitution $\hat{\Vect{u}} \rightarrow \Vect{y}$ yields Eq.~(\ref{eq12}).

\section{Derivation of Eq.~(\ref{eq15})}\label{BBB}

Substitution of Eq.~(\ref{eq12}) into Eq.~(\ref{eq14}) yields 
\begin{equation}\label{B_EQ_1}
    \begin{array}{l}
    \DS  \Big \langle \hat{\Vect{u}}_1\otimes \hat{\Vect{u}}_2\Big \rangle=\Bigg\langle \int_0^t \int_0^t \Bigl(e^{\Vect{A}_1(t-\tau_1)}\Vect{B}_1(\tau_1)\mathrm{d}\Vect{W}_1\Bigr) \otimes \Bigl(e^{\Vect{A}_2(t-\tau_2)}\Vect{B}_2(\tau_2)\mathrm{d}\Vect{W}_2\Bigr)\Bigg \rangle\\[4mm]
    \DS = \Bigg\langle\int_0^t \int_0^t  e^{\Vect{A}_1(t-\tau_1)}\Vect{B}_1(\tau_1)\mathrm{d}\Vect{W}_1 \otimes \Bigl( \Vect{B}_2(\tau_2)\mathrm{d}\Vect{W}_2\Bigr)  e^{\Vect{A}_2^\top(t-\tau_2)}\Bigg\rangle.
    \end{array}
\end{equation}
Here, the property of transposition of the matrix exponental~\cite{Hall}
\begin{equation}\label{B_EQ_2}(e^{\Vect{A}})^\top=e^{\Vect{A}^\top}.
\end{equation}
is used.
We write the expression for dyadic product of the vectors in Eq.~(\ref{B_EQ_1}) as
\begin{equation}\label{B_EQ_3}
\Vect{B}_1(t_1)\mathrm{d}\Vect{W}_1 \otimes \Bigl( \Vect{B}_2(t_2)\mathrm{d}\Vect{W}_2\Bigr) =\begin{pmatrix} 0 & 0  \\ 0 &  \mathrm{d}g_{1,2}\end{pmatrix},
\end{equation}
\begin{equation}\label{B_EQ_4}
\begin{array}{l}
\DS \mathrm{d}g_{1,2}\defeq \sum_{n,n'=0}^{\infty}b_n(t_1)b_{n'}(t_2)\cos{\frac{(2n+1)\theta_1}{2}}\cos{\frac{(2n'+1)\theta_2}{2}}\mathrm{d}W_{n,1}\mathrm{d}W_{n',2}\\[3mm]
\DS=\sqrt{\mathrm{d}t_1\mathrm{d}t_2}\sum_{n,n'=0}^{\infty}b_n(t_1)b_{n'}(t_2)\cos{\frac{(2n+1)\theta_1}{2}}\cos{\frac{(2n'+1)\theta_2}{2}}\rho_{n}(t_1)\rho_{n'}(t_2).
\end{array}
\end{equation}
By using the uncorrelatedness of the random variables~(see Sect.~\ref{sect2} and Eq.~(\ref{eq1})), we transform Eq.~(\ref{B_EQ_1}) to the form:
\begin{equation}\label{B_EQ_5}
    \DS  \Big \langle \hat{\Vect{u}}_1 \otimes  \hat{\Vect{u}}_2\Big \rangle= \int_0^t e^{\Vect{A}_1(t-\tau)}\Vect{G}(\tau)e^{\Vect{A}_2^\top(t-\tau)}\mathrm{d}\tau,\quad \Vect{G}(t)\defeq  \begin{pmatrix} 0 & 0  \\ 0 &  \DS\sum_{n=0}^\infty b_n^2(t)\cos{\frac{(2n+1)\theta_1}{2}}\cos{\frac{(2n+1)\theta_2}{2}}\end{pmatrix}.
\end{equation}
We see that the matrix~$\Vect{G}$ is equal to the product~$\Vect{B}_1\Vect{B}_2^\top$, \edited{which} proves Eq.~(\ref{eq15}). For a more rigorous derivation of the variance matrix, corresponding to the stochastic matrix equation in the form of Eq.~(\ref{eq10}), we recommend \edited{consulting} Sect.~6.4. in~\cite{Stepanov}.

\section{Details of numerics, related to Sect.~\ref{sect_4_4} }\label{CCC}
Below we provide the details of numerical simulations.
\par Consider the chain with~$N$ particles and two free ends. Following~\cite{Gavr2018}, introduce the following dimensionless values:
\begin{equation}\label{C_EQ_1}
\begin{array}{l}
    \DS \Tilde{u}\defeq u/a,\qquad \Tilde{v}\defeq v/v_s,\\[3mm]
    \DS \Tilde{b}\defeq b/(v_s\sqrt{\omega_e}),\qquad \Tilde{\eta}\defeq \eta/\omega_e,\\[3mm]
    \Tilde{t}\defeq \omega_et,\qquad \Tilde{T}\defeq k_\mathrm{B}T/(mv_s^2).
\end{array}
\end{equation}
Then, taking into account Eq.~(\ref{SUPPLY_KRONECK}), we rewrite Eqs.~(\ref{eq1}) in the form
\begin{equation}\label{C_EQ_2}
    \begin{array}{l}
\mathrm{d}\Tilde{u}_n=\Tilde{v}_n\mathrm{d}\Tilde{t},\quad n=0,1,...,N-1,\\[3mm] 
     \mathrm{d}\Tilde{v}_n=\left(F_n-2\Tilde{\eta} \Tilde{v}_n\right)\mathrm{d}\Tilde{t}+\Tilde{b}\delta_{n,j}\rho_j\sqrt{\mathrm{d}\Tilde{t}},\\[3mm] F_n(\Tilde{u}_n)\defeq (\Tilde{u}_{n+1}-\Tilde{u}_n)(1-\delta_{n,N-1})-(\Tilde{u}_n-\Tilde{u}_{n-1})(1-\delta_{n,0}),
    \end{array}
\end{equation}
with zero initial conditions with respect to the dimensionless particle displacement and velocity.  The kinetic temperature, $\Tilde{T}_n$, is calculated by its definition~(\ref{eq4}), where the mathematical
expectation is replaced by \edited{an} average over~$N_\mathrm{R}$~realizations:
\begin{equation}\label{C_EQ_3}
    \Tilde{T}_n=\frac{1}{N_\mathrm{R}}\sum_{k=1}^{N_\mathrm{R}}(\Tilde{v}_{n}\vert_k)^2,
\end{equation}
where~$\Tilde{v}_{n}\vert_k$ is the particle velocity for the~$k$-th~realization. To obtain the particle velocity, we integrate the Eqs.~(\ref{C_EQ_2}) using the method, proposed in~\cite{BAOAB}. The following step of the numerical simulation is performed in accordance with the following scheme, which consists of five sequentially implemented sub-steps:
\begin{equation}\label{C_EQ_4}
\begin{array}{l}
    \DS \Tilde{v}_{n}\left(\Tilde{t}+\frac{\Delta \Tilde{t}}{2}\right)=\Tilde{v}_n(\Tilde{t})+F_n\left(\Tilde{u}_n(\Tilde{t})\right)\frac{\Delta \Tilde{t}}{2},\\[4mm] \DS \Tilde{u}_{n}\left(\Tilde{t}+\frac{\Delta \Tilde{t}}{2}\right)=\Tilde{u}_{n}(\Tilde{t})+\Tilde{v}_n\left(\Tilde{t}+\frac{\Delta \Tilde{t}}{2}\right)\frac{\Delta \Tilde{t}}{2},\\[4mm]
    
    \DS \Tilde{v}_{n}\left(\Tilde{t}+\frac{\Delta \Tilde{t}}{2}\right)=\Tilde{v}_{n}\left(\Tilde{t}+\frac{\Delta \Tilde{t}}{2}\right)e^{-2\Tilde{\eta}\Delta\Tilde{t}}+\Tilde{b}\delta_{n,j}\rho_j\sqrt{\frac{1-e^{-4\Tilde{\eta}\Delta\Tilde{t}}}{4\Tilde{\eta}}},\\[5mm]

     \DS \Tilde{u}_{n}\left(\Tilde{t}+\Delta \Tilde{t}\right)=\Tilde{u}_{n}\left(\Tilde{t}+\frac{\Delta \Tilde{t}}{2}\right)+\edited{\Tilde{v}_{n}\left(\Tilde{t}+\frac{\Delta \Tilde{t}}{2}\right)}\frac{\Delta \Tilde{t}}{2},\\[4mm]
     \DS \Tilde{v}_{n}\left(\Tilde{t}+\Delta \Tilde{t}\right)=\edited{\Tilde{v}_{n}\left(\Tilde{t}+\frac{\Delta \Tilde{t}}{2}\right)}+F_n\left(\Tilde{u}_{n}\left(\Tilde{t}+\Delta \Tilde{t}\right)\right)\frac{\Delta \Tilde{t}}{2},
\end{array}
\end{equation}
where~$\Delta \Tilde{t}$ is time step.
\par \small \edited{\textit{Remark}~4. 
For~$\Tilde{\eta}=0$, the noise term~(Eq.~(\ref{C_EQ_4}), third line) has the form~$\Tilde{b}\delta_{n,j}\rho_j\sqrt{\Delta \Tilde{t}}$.}
\normalsize

In numerical simulations, the following parameters are used:
\begin{equation}\label{C_EQ_6}
    \Delta \Tilde{t}=0.01,\quad \Tilde{b}=1,\quad N=512,\quad N_\mathrm{R}=10^5.
\end{equation}
For each time step (and realization) a generated random number~$\rho_j$ is uniformly distributed on the segment~$[-\sqrt{3}; \sqrt{3}]$, \edited{which} satisfies the conditions imposed on uncorrelatedness of the random numbers~(see Sect.~\ref{sect2}). To speed up calculations, the library <<OpenMP>> is used. \edited{The chain length \(N=512\) is chosen so that, over the time intervals
used in Sect.~\ref{sect_4_4} the waves emitted from the heat-supply region do
not return from the remote free end to the observation region. The
time step and the number of realizations provide stable ensemble
averages for the curves shown in Figs.~\ref{fig9}, \ref{fig11}--\ref{stat_h0}.}
\end{appendices}

\end{document}